\documentclass[11pt,a4paper]{article}
\pdfoutput=1

\usepackage{jcappub}
\usepackage[utf8]{inputenc}
\usepackage{graphicx}
\usepackage{epsfig} 
\usepackage{subcaption}
\usepackage{color}
\usepackage{comment}
\usepackage{hyperref}

\def\beq{\begin{equation}}
\def\eeq{\end{equation}}
\def\baq{\begin{eqnarray}}
\def\eaq{\end{eqnarray}}

\newcommand{\be}{\begin{equation}} 
\newcommand{\ee}{\end{equation}}
\newcommand{\bea}{\begin{equation}\begin{aligned}} 
\newcommand{\eea}{\end{aligned}\end{equation}}

\newcommand{\bmp}{\noindent\begin{minipage}{16cm}}
\newcommand{\emp}{\end{minipage}\vskip 7mm} 
\def\lsim{\mathrel{\raise.3ex\hbox{$<$\kern-.75em\lower1ex\hbox{$\sim$}}}}
\def\gsim{\mathrel{\raise.3ex\hbox{$>$\kern-.75em\lower1ex\hbox{$\sim$}}}}

\newcommand{\intron}[1]{}

\title{Attractor Behaviour in Multifield Inflation}

\author[]{Pedro Carrilho,}
\author[]{David Mulryne,}
\author[]{John Ronayne, and}
\author[]{Tommi Tenkanen}

\affiliation[]{Astronomy Unit, Queen Mary University of London,
 \\ Mile End Road, London, E1 4NS, United Kingdom}

\emailAdd{p.gregoriocarrilho@qmul.ac.uk}
\emailAdd{d.mulryne@qmul.ac.uk}
\emailAdd{j.ronayne@qmul.ac.uk}		
\emailAdd{t.tenkanen@qmul.ac.uk}

\abstract{We study multifield inflation in scenarios where the fields are coupled non-minimally to gravity via $\xi_I(\phi^I)^n g^{\mu\nu}R_{\mu\nu}$, where $\xi_I$ are coupling constants, $\phi^I$ the fields driving inflation, $g_{\mu\nu}$ the space-time metric, $R_{\mu\nu}$ the Ricci tensor, and $n>0$. We consider the so-called $\alpha$-attractor models in two formulations of gravity: in the usual metric case where $R_{\mu\nu}=R_{\mu\nu}(g_{\mu\nu})$, and in the {\it Palatini} formulation where $R_{\mu\nu}$ is an independent variable. As the main result, we show that, regardless of the underlying theory of gravity, the field-space curvature in the Einstein frame has no influence on the inflationary dynamics at the limit of large $\xi_I$, and one effectively retains the single-field case. However, the gravity formulation does play an important role: in the metric case the result means that multifield models approach the single-field $\alpha$-attractor limit, whereas in the Palatini case the attractor behaviour is lost also in the case of multifield inflation. We discuss what this means for distinguishing between different models of inflation.}

\keywords{Non-minimal coupling to gravity, multifield inflation, $\alpha$-attractors, metric gravity, Palatini gravity}

\begin{document}
\maketitle


\section{Introduction}
\label{introduction}

The standard paradigm in modern cosmology is that a period of very rapid expansion, cosmic inflation, preceded the usual Hot Big Bang era. The key reason to assume that cosmic inflation really happened lies in explaining the origin of the small inhomogeneities observed in the Cosmic Microwave Background (CMB) radiation spectrum~\cite{Starobinsky:1980te, Sato:1980yn, Guth:1980zm, Linde:1981mu, Albrecht:1982wi, Linde:1983gd,Lyth:1998xn,Mazumdar:2010sa,Martin:2013tda,Patrignani:2016xqp}. 

Amongst the parameters of the curvature power spectrum that characterize these primordial fluctuations, the Planck satellite has measured two to a very high accuracy: the amplitude of the spectrum, $A_s=(2.141\pm 0.052)\times 10^{-9}$, and the corresponding spectral tilt, $n_s=0.9681\pm 0.0044$ (both at the $1\sigma$ level) \cite{Ade:2015xua}. On top of that, the Planck measurements have placed strong constraints on the amount of primordial non-Gaussianity \cite{Ade:2015ava} and primordial isocurvature perturbations \cite{Ade:2015lrj}. The BICEP2/Keck Array data has also recently placed an upper bound for the tensor-to-scalar ratio, $r<0.09$ (at the $2\sigma$ level) \cite{Array:2015xqh}. These are numbers that any successful model of inflation has to predict. However, the most important questions that remain are: why, how, and when did inflation happen? What are the microphysical models for inflation that best fit to data, and how to distinguish between them? 

In this paper, we concentrate on scenarios where the fields that take part in the inflationary dynamics couple non-minimally to the gravity sector of the theory. We study couplings of the type $\xi_I (\phi^I)^n g^{\mu\nu}R_{\mu\nu}$, where $\phi^I$ are a set of scalar fields that drive inflation, $\xi_I$ are coupling constants, $g_{\mu\nu}$ the space-time metric, and $R_{\mu\nu}$ the Ricci tensor. In addition we take $n>0$. This model class is of particular interest, as non-minimal couplings should be seen as a generic ingredient of coherent model frameworks, generated by quantum corrections in a curved space-time \cite{Birrell:1982ix}, and because such models typically predict values for inflationary observables close to the best fit to data \cite{Martin:2013tda,Ade:2015lrj}. In particular, this is the case for the scenario where the Standard Model (SM) Higgs is the inflaton field \cite{Bezrukov:2007ep,Bauer:2008zj}.

In Ref.~\cite{Kallosh:2013tua}, it was found that for
large values of the non-minimal coupling strength, all such single-field models 
asymptote to a universal attractor, the $R^2$ or Starobinsky model \cite{Starobinsky:1980te},
independently of the original scalar potential. In Ref.~\cite{Kaiser:2013sna} a similar conclusion was reached for a 
broad range of multifield models. Despite their simplicity, the predictions of these models fall perfectly into the region constrained by Planck and BICEP2/Keck Array data.
Such models where later on named as $\alpha$-attractor models, after which they have been studied in a large number of works, see e.g. \cite{Kallosh:2013maa,Kallosh:2013tua,Galante:2014ifa,Jarv:2016sow}. 

Recently, however, in Ref.~\cite{Jarv:2017azx} it was shown that $\alpha$-attractors are in fact not universal but depend on the underlying theory of gravity in a subtle way. The non-minimal couplings of the type $\xi_I (\phi^I)^n g^{\mu\nu}R_{\mu\nu}$ contain freedom to choose the space-time connection: one can either study the usual metric case where $R_{\mu\nu}=R_{\mu\nu}(g_{\mu\nu})$, or choose an alternative approach, the so-called {\it Palatini} formulation of gravity, where the connection $\Gamma$ and hence also $R_{\mu\nu}=R_{\mu\nu}(\Gamma)$ are independent variables. In the context of general theory of relativity, the metric formalism coincides with the one of Palatini, as minimising the Einstein-Hilbert action with respect to the connection uniquely fixes it to be of the Levi-Civita form, $\Gamma=\Gamma(g_{\mu\nu})$. In more general models, however, especially in the ones involving matter fields that are non-minimally coupled to gravity, these two formalisms lead to two inherently different gravitational theories~\cite{Sotiriou:2008rp,Clifton:2011jh,Nojiri:2017ncd}. This means that inflationary models with non-minimal couplings to gravity cannot be characterized just by the inflaton field potential, but that the connection must also be specified. This was originally studied in \cite{Bauer:2008zj,Tamanini:2010uq,Bauer:2010jg}, and has recently gained increasing attention, see \cite{Rasanen:2017ivk,Tenkanen:2017jih,Fu:2017iqg,Racioppi:2017spw,Markkanen:2017tun,Jarv:2017azx,Racioppi:2018zoy,Enckell:2018kkc,Garcia-Saenz:2018ifx}.

The novelty of our work lies in generalizing the analysis of non-minimally coupled models in both of the above formulations of gravity to the case of multifield inflation, where there is more than one field taking part in inflationary dynamics, and to study  attractor behaviour in this case. For other recent works on non-minimally coupled multifield inflation see e.g. \cite{Kaiser:2010ps,Kaiser:2015usz,Schutz:2013fua,Kaiser:2013sna,Kaiser:2012ak,White:2012ya,Kaiser:2010yu,Kallosh:2013maa}, and for previous studies on $\alpha$-attractors in the case of multiple fields, see \cite{Kallosh:2013daa,Linde:2016uec,Achucarro:2017ing,Christodoulidis:2018qdw}.
As the main result, we show that in the case of multiple, non-minimally coupled fields, the curvature of the field-space in the Einstein frame (where the non-minimal couplings vanish) has no influence on the inflationary dynamics at the limit of large $\xi_I$ in all the cases we study, and one effectively retains the single-field case in both of the above formulations of gravity. In the metric case this means that multifield models approach the single-field $\alpha$-attractor limit as seen in Ref.~\cite{Kaiser:2013sna}, whereas in the Palatini case the attractor behaviour is lost also in the case of multifield inflation. We will also discuss what this means for distinguishing between different models of inflation.  

The paper is organized as follows: in Section \ref{inflation}, we present the multifield models we are considering and perform the conformal transformation to the Einstein frame where the non-minimal couplings vanish. In Section \ref{results}, we present the numerical set-up and the results, discussing observational ramifications and demonstrating why the curved field-space metric in the Einstein frame has no influence on the inflationary dynamics. Finally, in Section \ref{conclusions}, we conclude.

\section{Multifield inflation with non-minimal couplings to gravity}
\label{inflation}

We consider a theory with multiple scalar fields, all of which are non-minimally coupled to gravity. This is explicit in the so-called Jordan frame action, which in our case is
\be
\label{nonminimal_action}
S_J = \int d^4x \sqrt{-g}\left(\frac{1}{2} \delta_{IJ}g^{\mu\nu}\partial_{\mu}\phi^I\partial_{\nu}\phi^J -\frac{M_{\rm P}^2}{2}\left(1 + f(\phi^I)\right) g^{\mu\nu}R_{\mu\nu}(\Gamma) - V(\phi^I)\right) ,
\ee
where $M_{\rm P}$ is the reduced Planck mass,
$g$ is the determinant of the metric, $\Gamma$ is the connection, and the Einstein summation convention has been employed both in the spacetime indices labelled by Greek letters ($\mu,\nu$) and in the field-space indices labelled by capital letters ($I,J$), for which the sum runs over the total number of fields. The potential $V(\phi^I)$ is at this point completely general and could, in principle, contain all possible mass and interaction terms of the scalar fields allowed by the underlying symmetries of the theory. The non-minimal coupling function $f(\phi^I)$ is also unspecified in the action, but will, in the following, generally take the form
\be
\label{nmc_function}
f(\phi^I)=\sum_I{\xi_I^{(n)}\left(\frac{\phi^I}{M_{\rm P}}\right)^n}\,,
\ee
with $\xi_I^{(n)}$ the dimensionless non-minimal coupling parameters. The most well studied of these couplings is the one generated by quantum corrections of a quartic scalar theory in a curved spacetime, for which $n=2$. For example, this is the case for the usual (single-field) Higgs inflation \cite{Bezrukov:2007ep}.

In the metric formulation of gravity, the connection $\Gamma$ is determined uniquely as a function of the metric tensor, i.e. it is $\bar{\Gamma}=\bar{\Gamma}(g_{\mu\nu})$ with
\be
	\bar{\Gamma}^\lambda_{\alpha\beta} = \frac{1}{2}g^{\lambda\rho}(\partial_\alpha g_{\beta\rho} + \partial_\beta g_{\rho\alpha} - \partial_\rho g_{\alpha\beta}) \,,
\ee
the Levi-Civita connection. In the Palatini formalism both $g_{\mu\nu}$ and $\Gamma$ are treated as independent variables, and the only assumption is that the connection is torsion-free, $\Gamma^\lambda_{\alpha\beta}=\Gamma^\lambda_{\beta\alpha}$. The application of the variational principle then gives rise to an extra equation for the connection, in addition to the one for the metric. For the Einstein-Hilbert action, the extra equation forces the connection to have the usual Levi-Civita form, but in more general theories of gravity, such as $f(R)$ theories, or in the presence of non-minimal couplings, this is no longer true in the Jordan frame.
 	
However, the non-minimal couplings in the Jordan frame action \eqref{nonminimal_action} can be removed by a conformal transformation to the Einstein frame, 
\begin{equation}
\label{Omega}
g_{\mu\nu} \to \Omega^{-1}(\phi^I)g_{\mu\nu}, \hspace{.5cm} \Omega(\phi^I)\equiv 1+f(\phi^I) \, .
\end{equation}	
After this transformation, the action \eqref{nonminimal_action} becomes
\be
S_{\rm E} = \int d^4x \sqrt{-g}\bigg(\frac{1}{2}G_{IJ}(\phi^I){\partial}_{\mu}\phi^I{\partial}^{\mu}\phi^J -\frac{1}{2}M_{\rm P}^2R - V(\phi^I)\Omega^{-2}(\phi^I)  \bigg),
\label{EframeS1}
\ee
where $R = g^{\mu\nu}R_{\mu\nu}(\bar{\Gamma})$, i.e. in the Einstein frame we retain the standard Levi-Civita connection regardless of the chosen theory of gravity, and the scalars have acquired a non-trivial field-space metric, given by
\be
\label{fieldmetric}
G_{IJ}=\Omega^{-1}\delta_{IJ}+\frac32 \upsilon M_{\rm P}^2\Omega^{-2}\frac{\partial\Omega}{\partial \phi^I}\frac{\partial\Omega}{\partial \phi^J}\,,
\ee
where $\upsilon=1$ in the metric case and $\upsilon=0$ in the Palatini case. With this conformal transformation, we have therefore transferred the dependence on the choice of gravitational degrees of freedom from the connection to the field-space metric. 

In the following, we will analyse inflation in both cases, metric and Palatini. For simplicity, we study two-field models with the potential
\be
\label{potential}
V(\phi,\sigma) = \lambda_\phi^{(2n)}  M_{\rm P}^{4-2n} \phi^{2n} + \lambda_\sigma^{(2n)}  M_{\rm P}^{4-2n} \sigma^{2n},
\ee
where $n>0$, $\lambda_\phi^{(2n)}$ and $\lambda_\sigma^{(2n)}$ are dimensionless coupling constants, and $M_{\rm P}^{4-2n}$ has been introduced to have a scalar potential with a mass dimension equal to four. Later on, in Sec. \ref{3fieldcase}, we will also discuss the case where more than two fields take part in inflationary dynamics.

In metric gravity, the above models are cosmological attractors, i.e. their predictions for observables asymptote to those of $R^2$ or Starobinsky inflation in the limit of strong non-minimal coupling $\xi$, see Eq. \eqref{starobinsky_nsr}. This is, however, known not to be true for the single-field case in the Palatini scenario~\cite{Jarv:2017azx}, and in this paper we will test it also in a multifield case.

For the potential \eqref{potential}, the Einstein frame potential is
\be
U(\phi,\sigma) = \Omega(\phi,\sigma)^{-2} V(\phi,\sigma)= \frac{\lambda_\phi^{(2n)}  M_{\rm P}^{4-2n} \phi^{2n} + \lambda_\sigma^{(2n)}  M_{\rm P}^{4-2n} \sigma^{2n}}{\left(1+\xi_\phi^{(n)}\left(\frac{\phi}{M_{\rm P}}\right)^n+\xi_\sigma^{(n)} \left(\frac{\sigma}{M_{\rm P}}\right)^n\right)^2}\,.
\ee

For this and all other models in this formulation, the potential $U$ is the same for both metric and Palatini gravity. The major difference between the two is the Einstein frame field-space metric, $G_{IJ}$. We will therefore focus mostly on the parameters appearing in $G_{IJ}$ in our analysis, namely the non-minimal couplings, $\xi_I^{(n)}$. The overall amplitude of the parameters $\lambda_I^{(2n)}$ in Eq. \eqref{potential} can be fixed by requiring that the curvature power spectrum, defined as
\be
\label{powerspec}
\left\langle\zeta(\vec k_1) \zeta(\vec k_2) \right\rangle= (2\pi)^3\delta^{(3)}(\vec k_1+\vec k_2) P_\zeta(k_1)\,,\ \ \  \mathcal{P}_{\zeta}(k)=\frac{k^3}{2\pi^2}P_\zeta(k)\,,
\ee
has the measured amplitude,  $\mathcal{P}_{\zeta}=(2.141\pm 0.052)\times 10^{-9}$ (at the $68\%$ confidence level) \cite{Ade:2015xua}. Their ratio, however, is unconstrained and does play a role in the dynamics, as we will show.

In the following, we calculate the predictions for observables in this type of models. We compute the usual spectral index of curvature perturbations, $n_s$, defined by
\begin{equation}
\label{nsdef}
n_s-1=\frac{d \log \mathcal{P}_\zeta}{d \log k}\,;
\end{equation}
the tensor-to-scalar ratio, $r$, given by the ratio of the tensor and scalar power spectra
\begin{equation}
\label{rdef}
r=\frac{P_T}{P_\zeta}\,,
\end{equation}
with $P_T$ the tensor power spectrum, defined in the same way as $P_\zeta$, via Eq.~\eqref{powerspec}; and the amount of non-Gaussianity, measured via the amplitude of the reduced bispectrum in the equilateral configuration
\begin{equation}
\label{fnldef}
f_{\rm NL}=\frac{5}{18}\frac{B_\zeta(k,k,k)}{P_\zeta(k)^2}\,,
\end{equation}
in which $B_\zeta(k_1,k_2,k_3)$ is the bispectrum, defined via
\be
\label{bispec}
\left\langle\zeta(\vec k_1) \zeta(\vec k_2) \zeta(\vec k_3) \right\rangle= (2\pi)^3\delta^{(3)}(\vec k_1+\vec k_2+\vec k_3) B_\zeta(k_1, k_2, k_3)\,.
\ee
All of the above variables are evaluated at horizon crossing of the Planck pivot scale, $k=0.05\, {\rm Mpc}^{-1}$, which we take to correspond to modes which crossed the horizon $60$ e-folds before the end of inflation. We explore the parameter space of the models under consideration by varying all parameters of the scalar potential and the field-space metric, as well as the initial conditions for the evolution during inflation. In order to compute the predictions, we employ the transport method \cite{Dias:2016rjq} (see Refs.~\cite{Mulryne:2013uka,Anderson:2012em,Seery:2012vj,Mulryne:2010rp,Mulryne:2009kh,Dias:2011xy,Dias:2014msa,Dias:2015rca} for earlier related work) and the open source PyTransport code\footnote{The package is available at \href{https://github.com/jronayne/PyTransport}{github.com/jronayne/PyTransport}.}~\cite{DavidJohn1}.  The results and the set-up for finding initial conditions are presented in the next section. The transport approach 
evolves the two and three-point function of 
field fluctuations from initial conditions set in the quantum regime on sub-horizon scales (as well as the two point function of tensor perturbations), and includes all tree-level contributions. It then uses these correlations to calculate the power spectrum and bispectrum of $\zeta$. It was recently extended to include a non-trivial field-space metric in Refs.~\cite{DavidJohn2,Butchers:2018hds} (and is also the basis of another open source package CppTransport \cite{Seery:2016lko}).  


\section{Results}
\label{results}

\subsection{Numerical Set-up}

For a given sets of model parameters, we explore the initial condition space by first calculating an approximate position in field-space corresponding to 73 e-folds before the end of inflation\footnote{The number $N=73$ is chosen to start the evolution so that the modes which cross the horizon 60 e-folds before the end of inflation are accurately evolved in the sub-horizon stage.}. Before sampling, we transform our fields to polar form. Then we sample an angle from a uniform distribution. Following that we incrementally increase the radial distance from the minimum until a coordinate in field space is found for which inflation lasts 73 e-folds. Sampling over the full distribution of angles would reveal an approximate 73 e-fold surface in the field space. Next we transform our fields back to their Cartesian form and numerically evolve the background equations forward in time until the end of inflation. This provides a set of evolutions of roughly $73$ e-folds. For each set of model parameters the process is repeated with a new random angle. Finally, we evaluate the observables of interest -- $n_s$, $r$ and $f_{\rm NL}$ as defined in Eqs. \eqref{nsdef}--\eqref{fnldef} -- at the scale which left the horizon 60 e-folds before the end of inflation. We repeat this procedure for a representative set of values of the model parameters focusing mostly on the effect of the non-minimal couplings, $\xi_I$.

\begin{figure}[!htb]
	\centering
	        \includegraphics[width=0.45\textwidth]{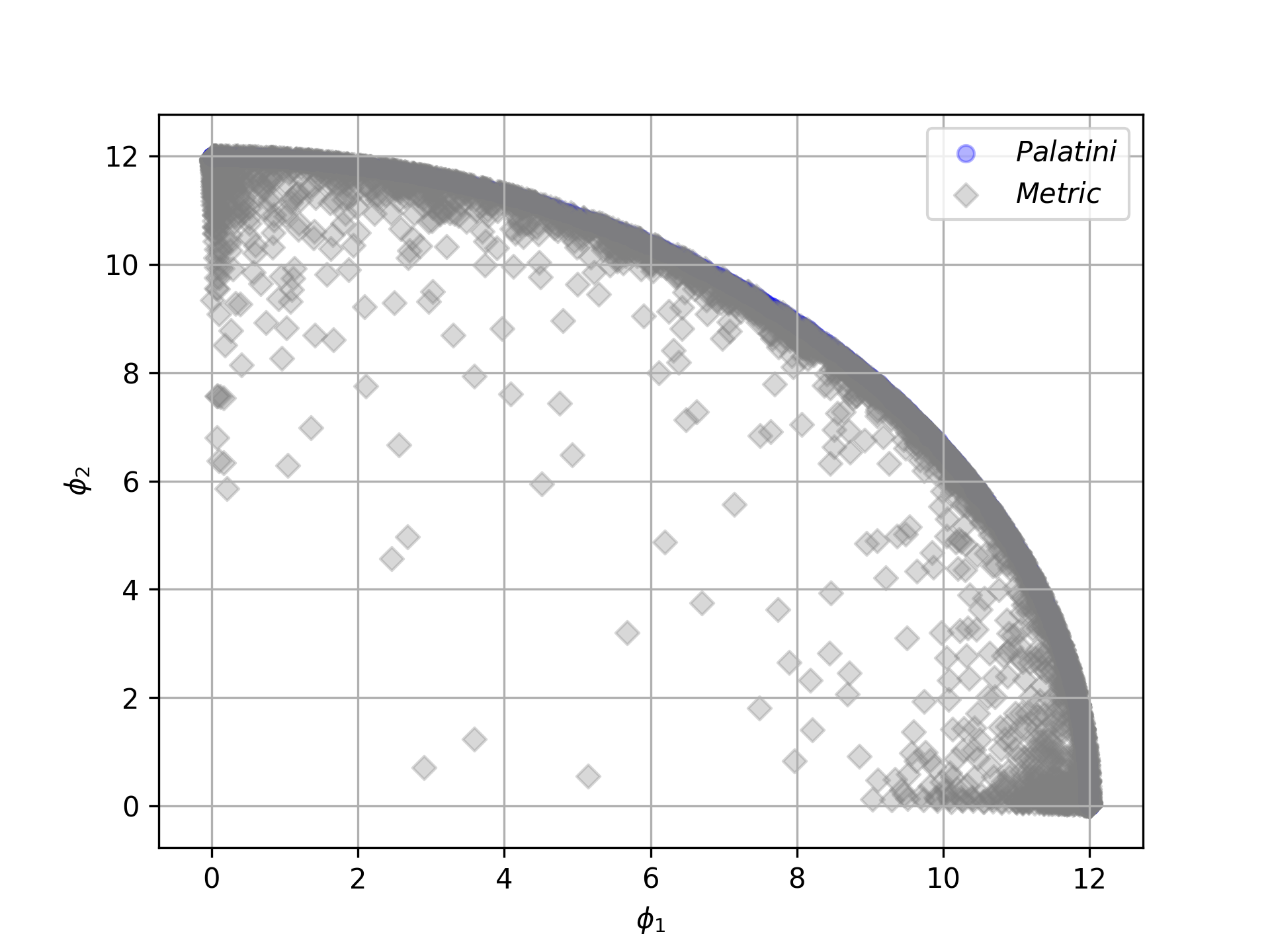}
	        \includegraphics[width=0.45\textwidth]{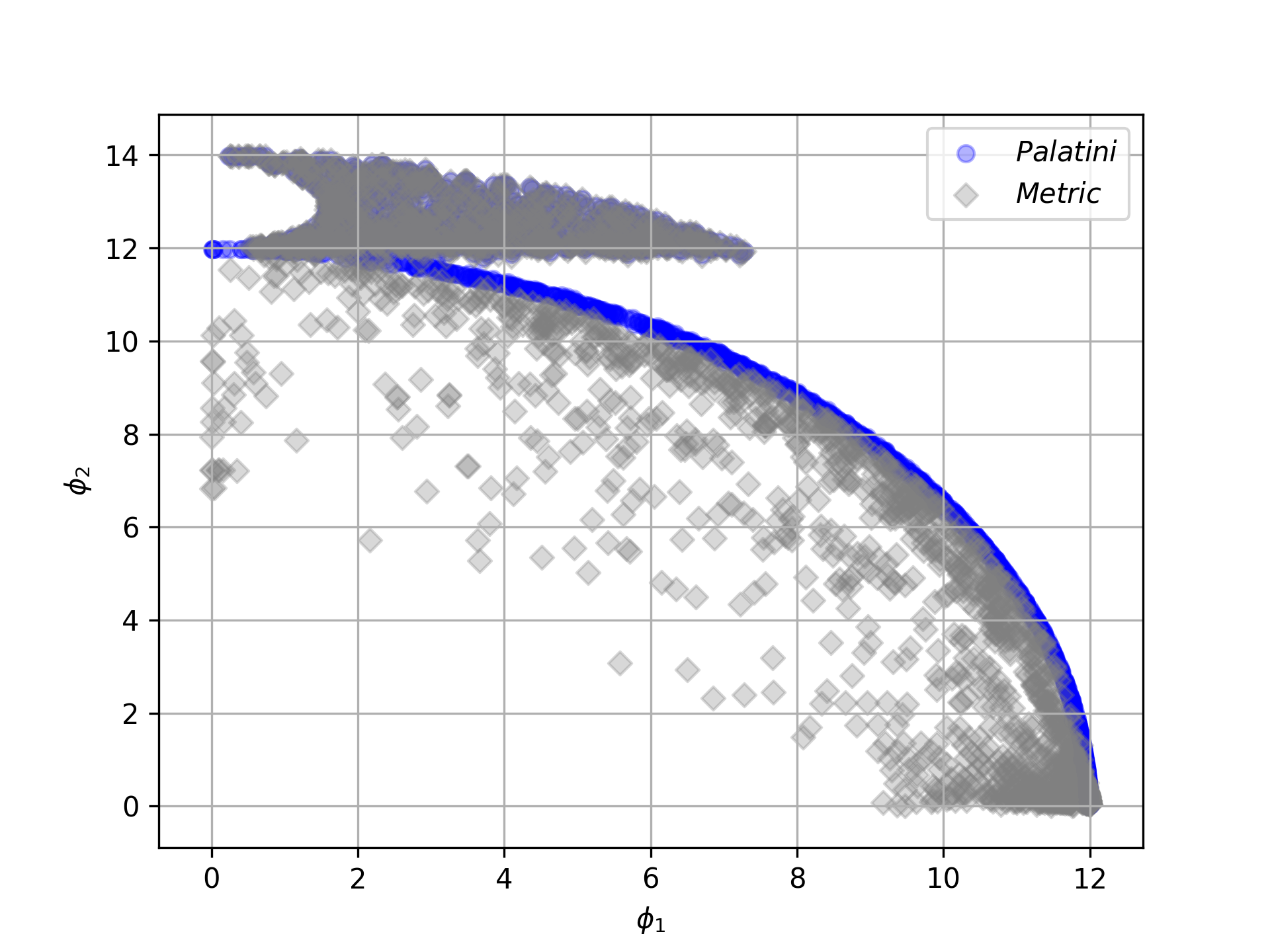}
        	\includegraphics[width=0.45\textwidth]{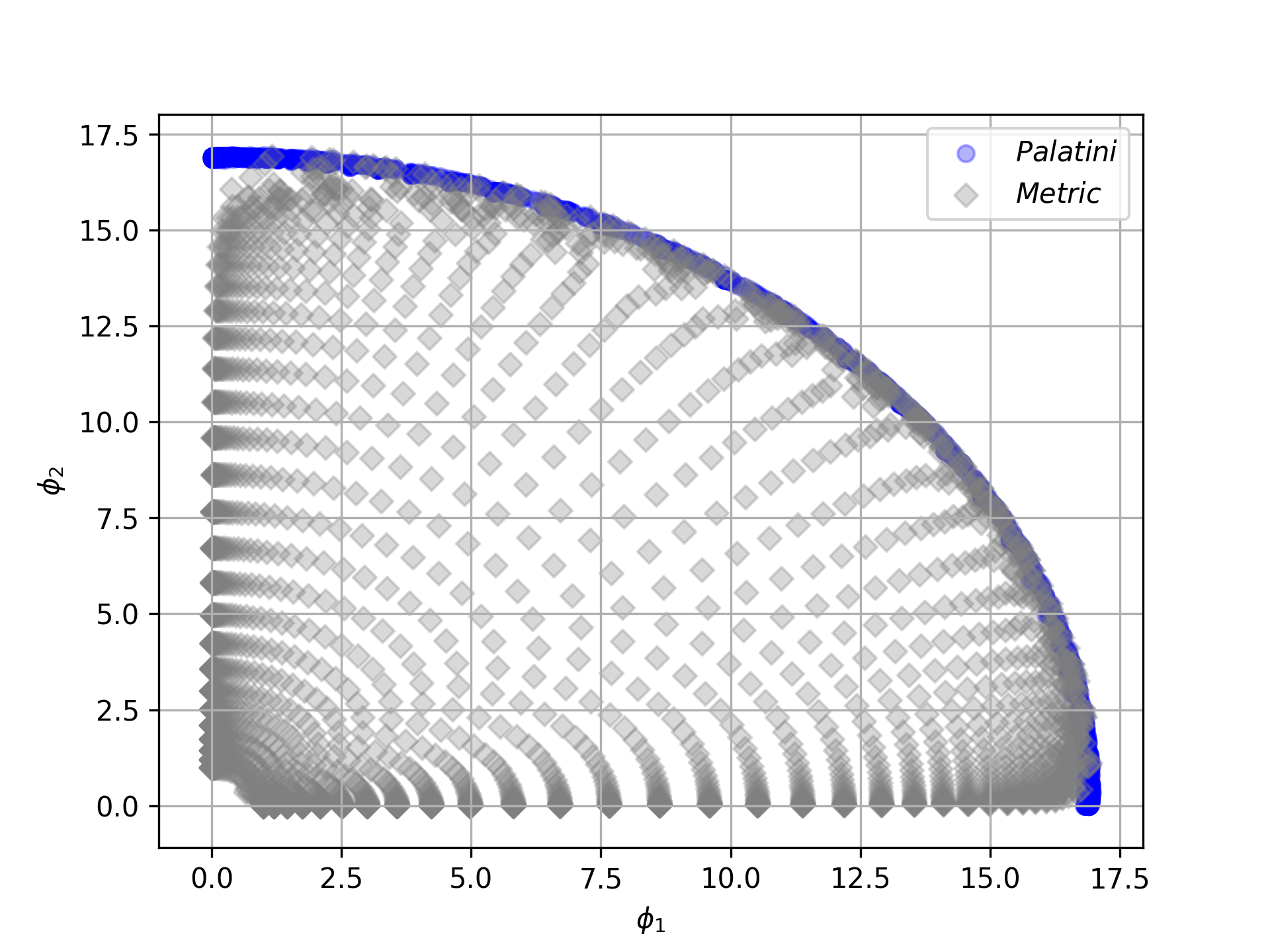}
        	\includegraphics[width=0.45\textwidth]{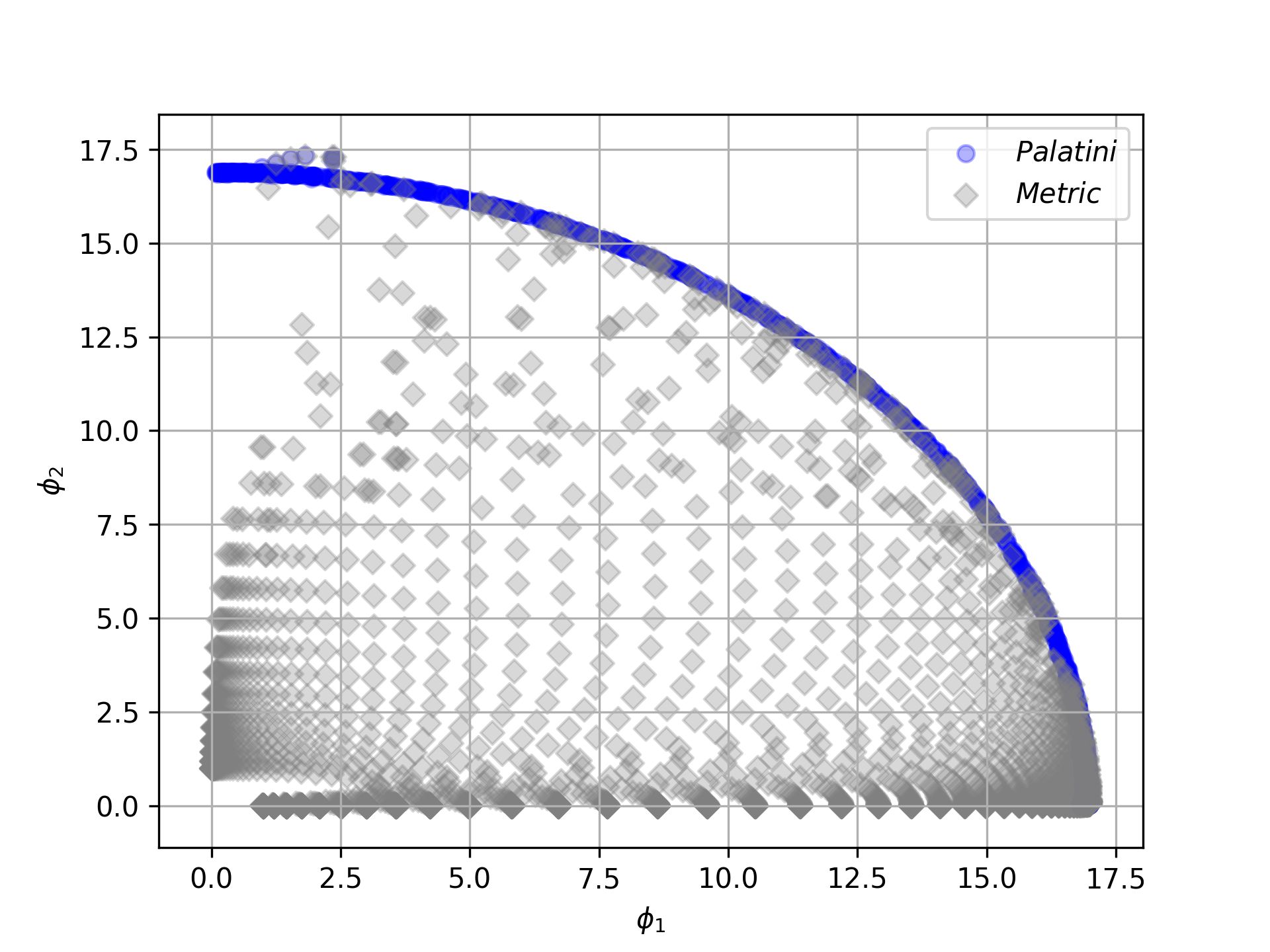}
        	\includegraphics[width=0.45\textwidth]{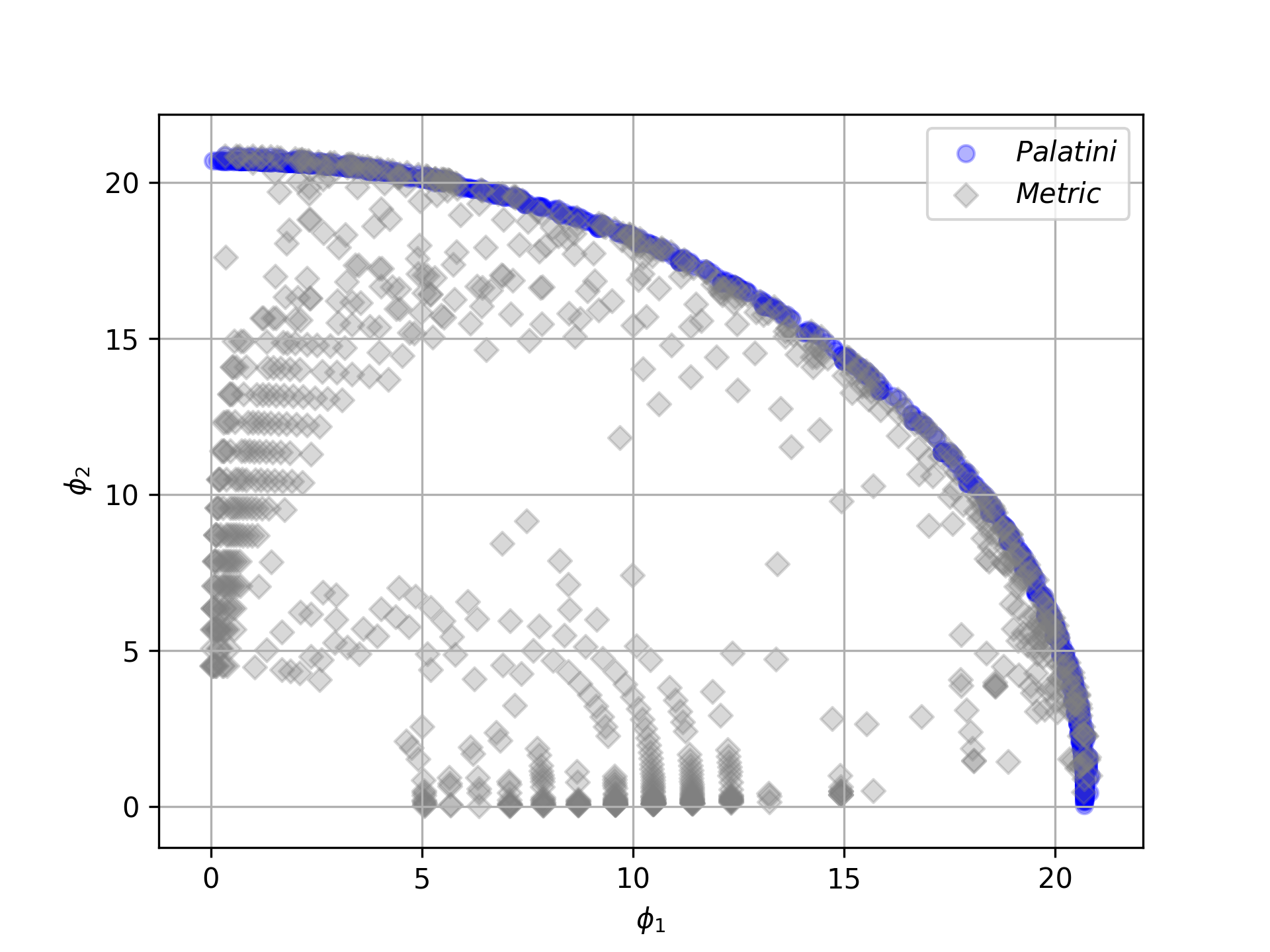}
 	        \includegraphics[width=0.45\textwidth]{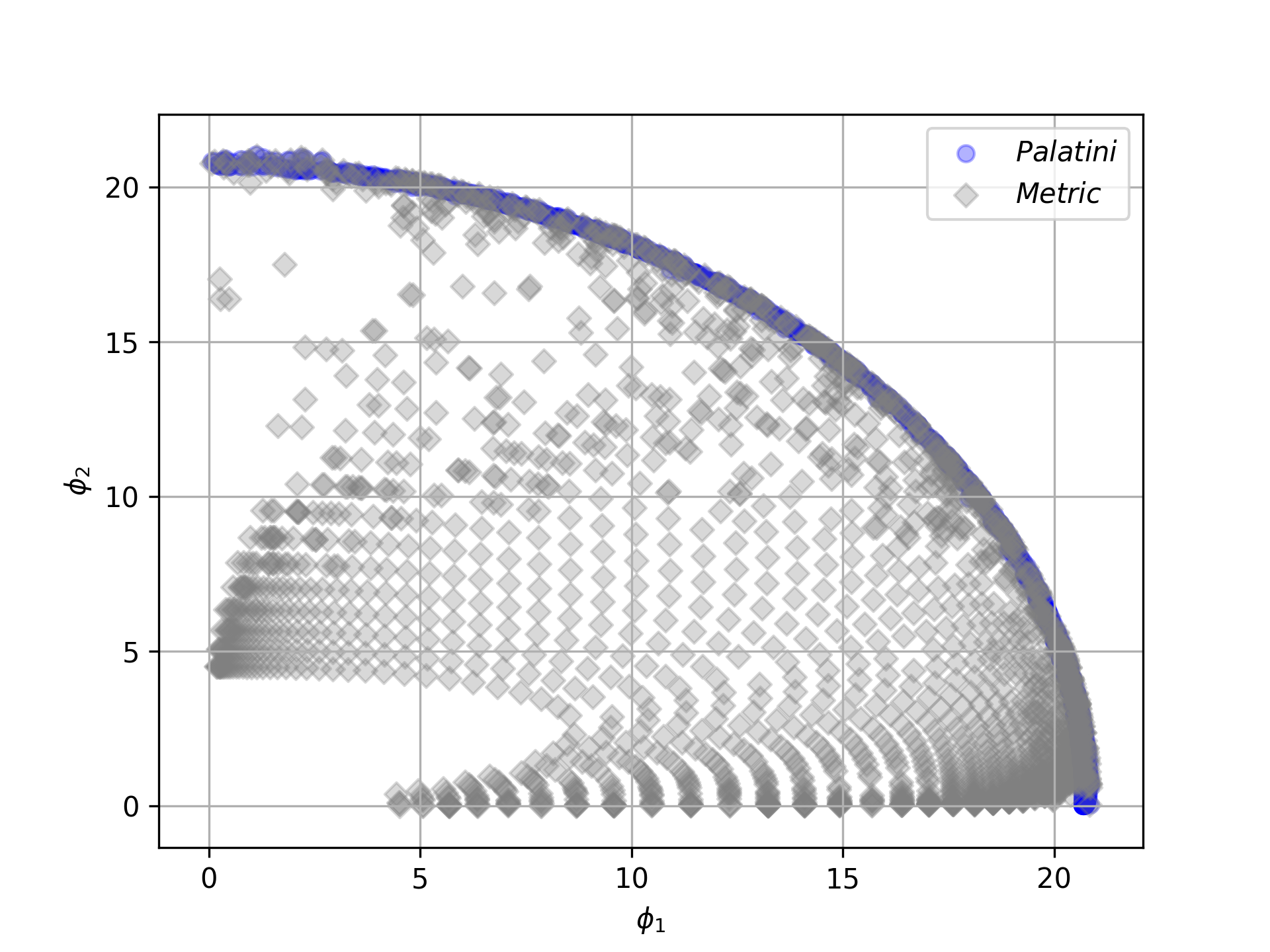}
        	\includegraphics[width=0.45\textwidth]{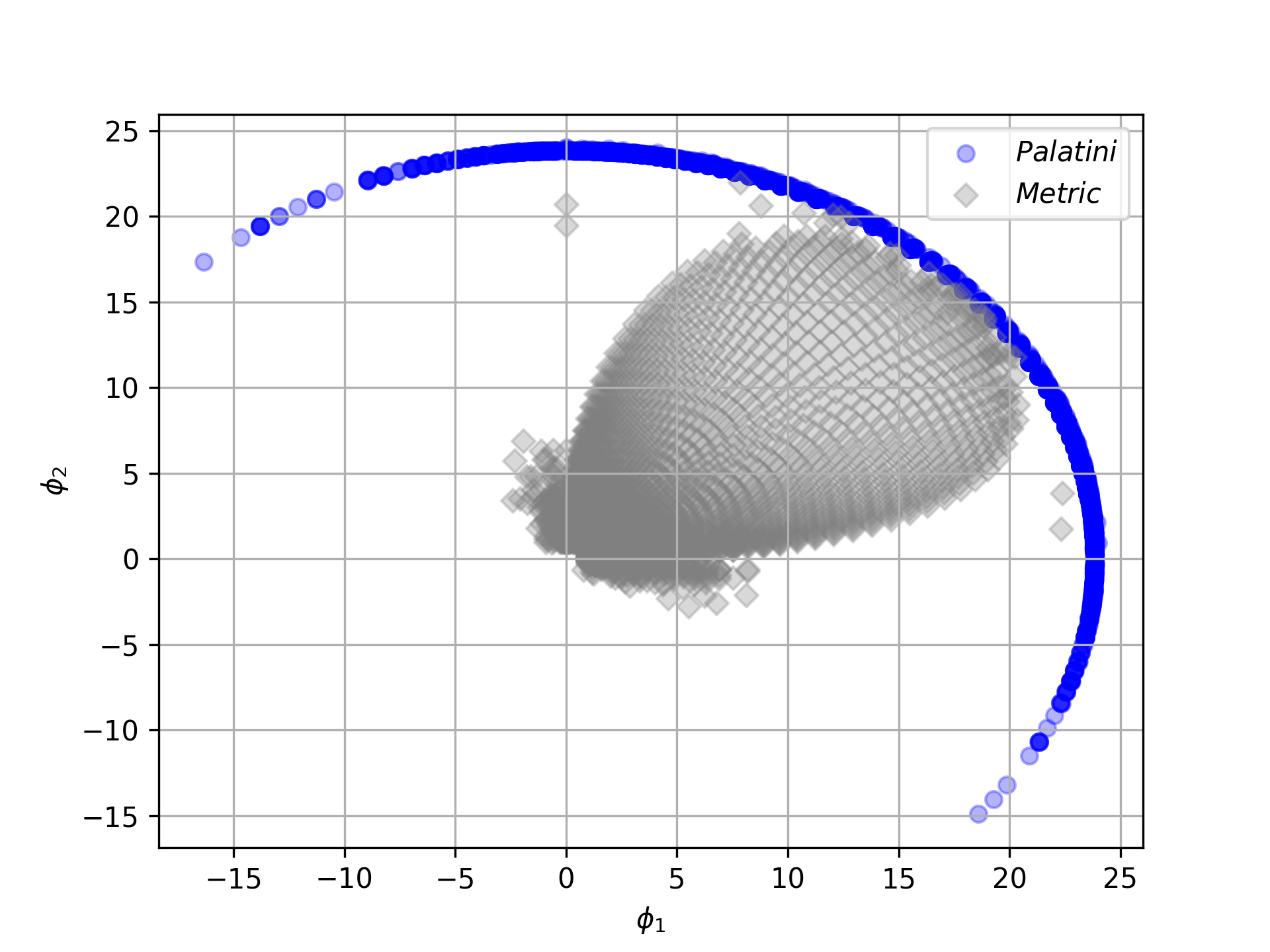}
        	\includegraphics[width=0.45\textwidth]{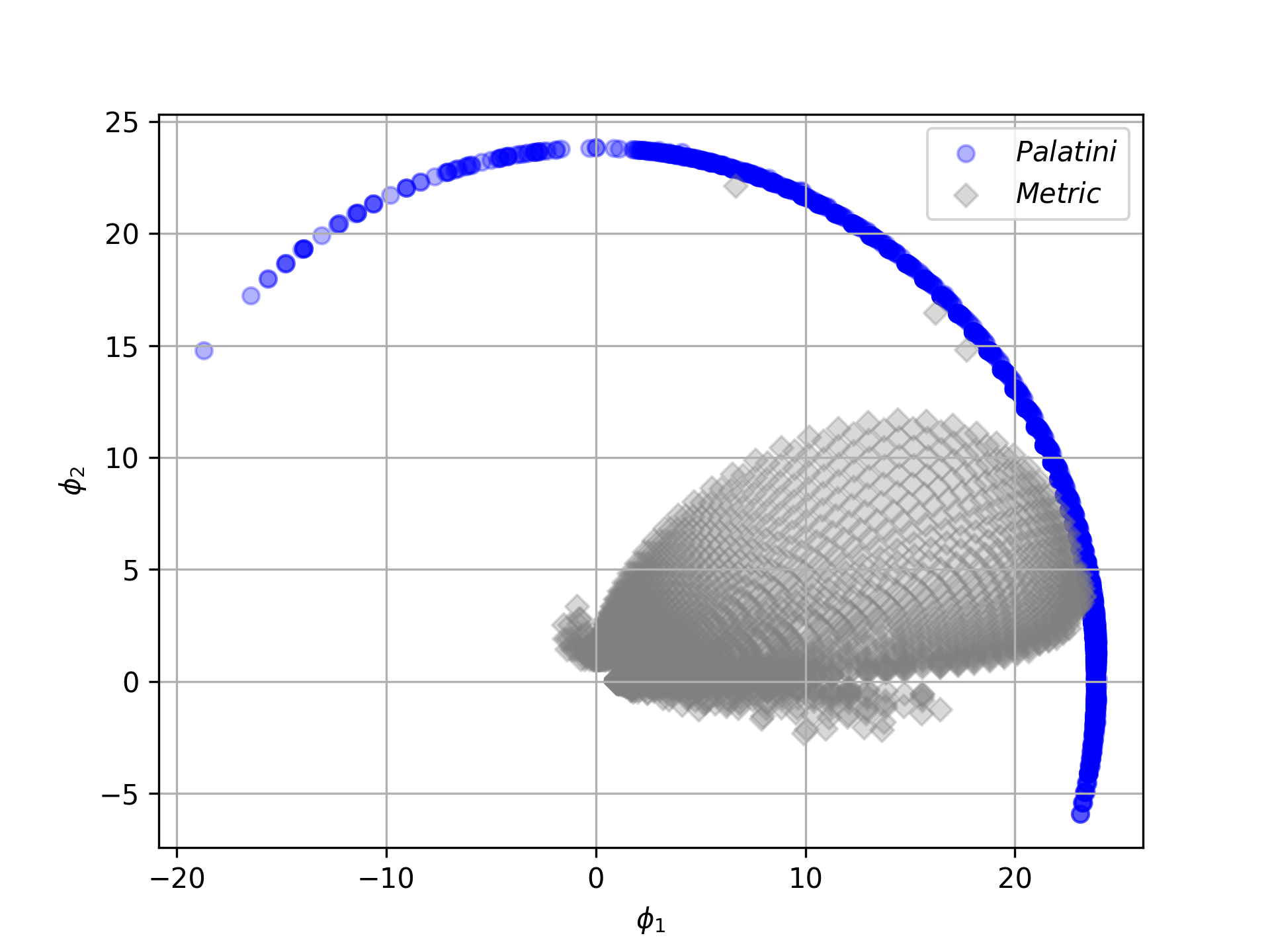}
	\caption{Sampling of initial conditions for metric (grey) and Palatini gravity (blue), $n=(1/2,1,3/2,2)$ from top to bottom. The left and right panels show the scenarios for different parameter ratios: $\lambda_\sigma/\lambda_\phi=19/14$ (left) and $\lambda_\sigma/\lambda_\phi=95/14$ (right). In all cases $\xi$ is varied between $(10^{-3}, 10)$.}\label{fig:ICs}
\end{figure}

Already at the background level, the evolution is different between metric and Palatini gravity. We can clearly see this in Fig.~\ref{fig:ICs}, which shows the initial conditions corresponding to 73 e-folds of inflation for both metric and Palatini gravity, with varying strengths of the non-minimal couplings. For Palatini gravity, the initial condition surface is independent of the value of the non-minimal coupling for nearly all cases, while for metric gravity the distance from the origin decreases with $\xi_I$ regardless of the value of $n$.

One can understand this by using the slow-roll approximation where inflation is sustained while the slow-roll parameter $\epsilon \equiv -\dot{H}/H^2$ and it's time derivative $\eta\equiv\dot{\epsilon}/\left(H\epsilon\right)$ are small for a sufficiently long period. We also assume that the background trajectories are approximately radial. Writing the fields in polar coordinates as
\begin{equation}
\phi=\rho\cos \psi \,,\ \ \sigma=\rho\sin\psi\,,
\end{equation}
the number of e-folds can be approximated by
\begin{equation}
N\approx \int_{\rho_e}^{\rho_i} \frac{U}{U_{,\rho}}G_{\rho\rho}\text{d}\rho\,,
\end{equation}
in which we use the notation $U_{,\rho}$ for a derivative in the direction of the radial coordinate $\rho$. All of the quantities in the integrand above can be calculated straightforwardly, given the field-space metric and the Einstein frame potential. To further simplify the notation, we also write the non-minimal couplings in polar coordinates as
\begin{align}
\xi_\phi=\xi \cos \theta\,,\ \ \xi_\sigma=\xi \sin \theta\,.\label{polarxi}
\end{align}
For Palatini gravity, the result is independent of $\xi$ and given by
\begin{equation}
N\approx \frac{\rho_\text{i}^2-\rho_\text{e}^2}{4nM_{\rm P}^2}\,,
\end{equation}
in which $\rho_\text{i}$ is the value of $\rho$ when the mode of interest exists the horizon and $\rho_\text{e}$ is the value at the end of inflation. Interestingly, this is exactly the same result as for $\xi=0$, which is why the initial conditions for the Palatini case coincide with those for the metric case at low $\xi$. For metric gravity and large values of $\xi$, the leading order term in the expansion in $\xi^{-1}$ is
\begin{equation}
N\approx \xi F_n(\psi,\theta)\frac{\rho_\text{i}^n-\rho_\text{e}^n}{M_{\rm P}^n}\,,
\end{equation}
which shows that to keep the number of e-folds constant, one requires smaller $\rho_\text{i}$ for larger $\xi$, as indeed is the case in Fig.~\ref{fig:ICs}. The function $F_n(\psi,\theta)$ simplifies to the single-field result when $\psi=\theta=0$ or $\psi=\theta=\pi/2$, which, for $n=2$, is $F_2(0,0)=3/4$, matching the result in Ref.~\cite{Bauer:2008zj}.

We see that this approximation works generically very well, except when the parameter ratio is large in certain directions in the field-space. This is because the approximation of radial trajectories fails in those cases, rendering the above approximate result inapplicable. This emphasizes the importance of accurate numerical analysis of multifield models, to which we now turn.

\subsection{Attractor models}

\begin{figure}
	\centering
		\includegraphics[width=0.45\textwidth]{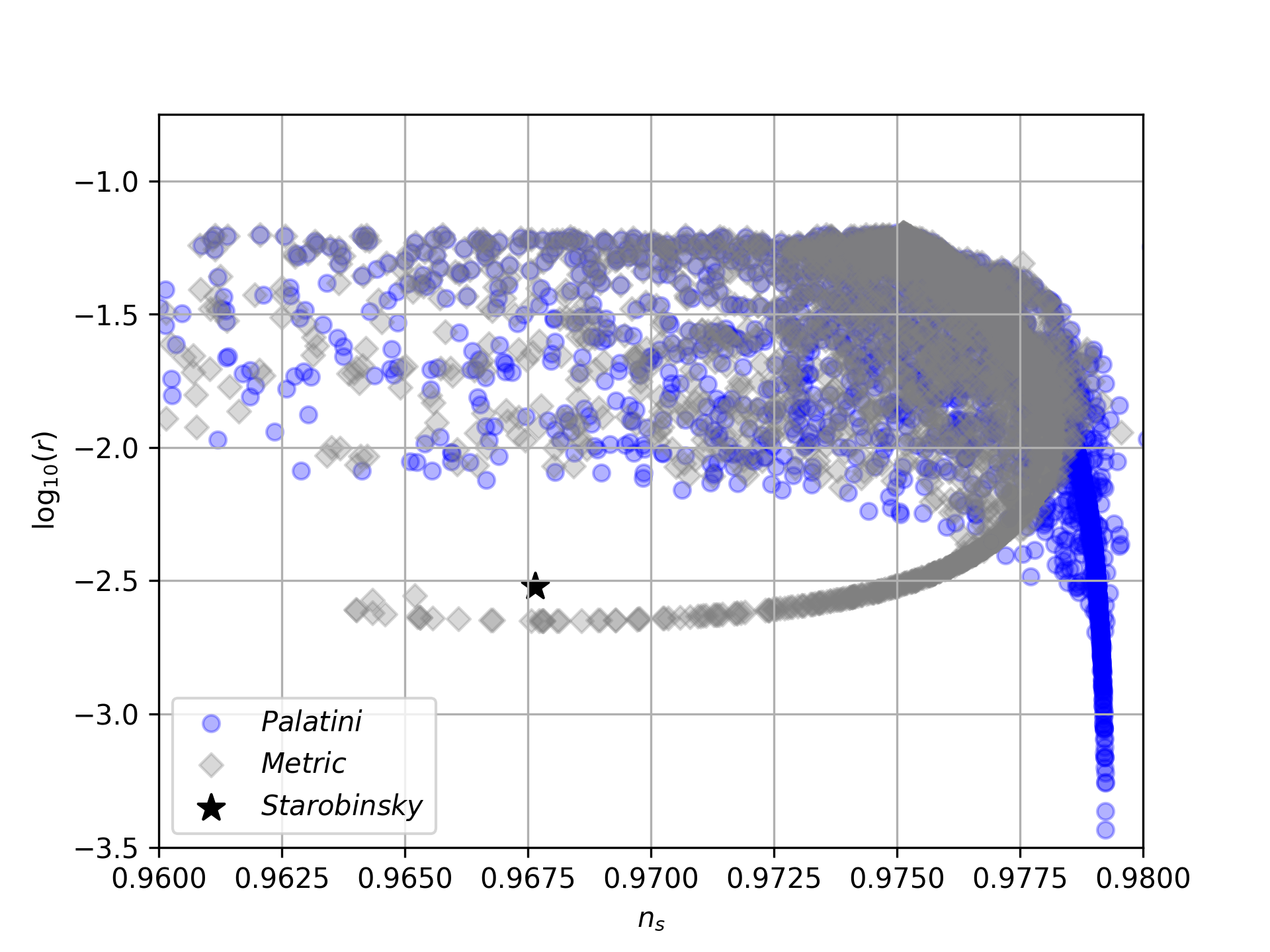}
        	\includegraphics[width=0.45\textwidth]{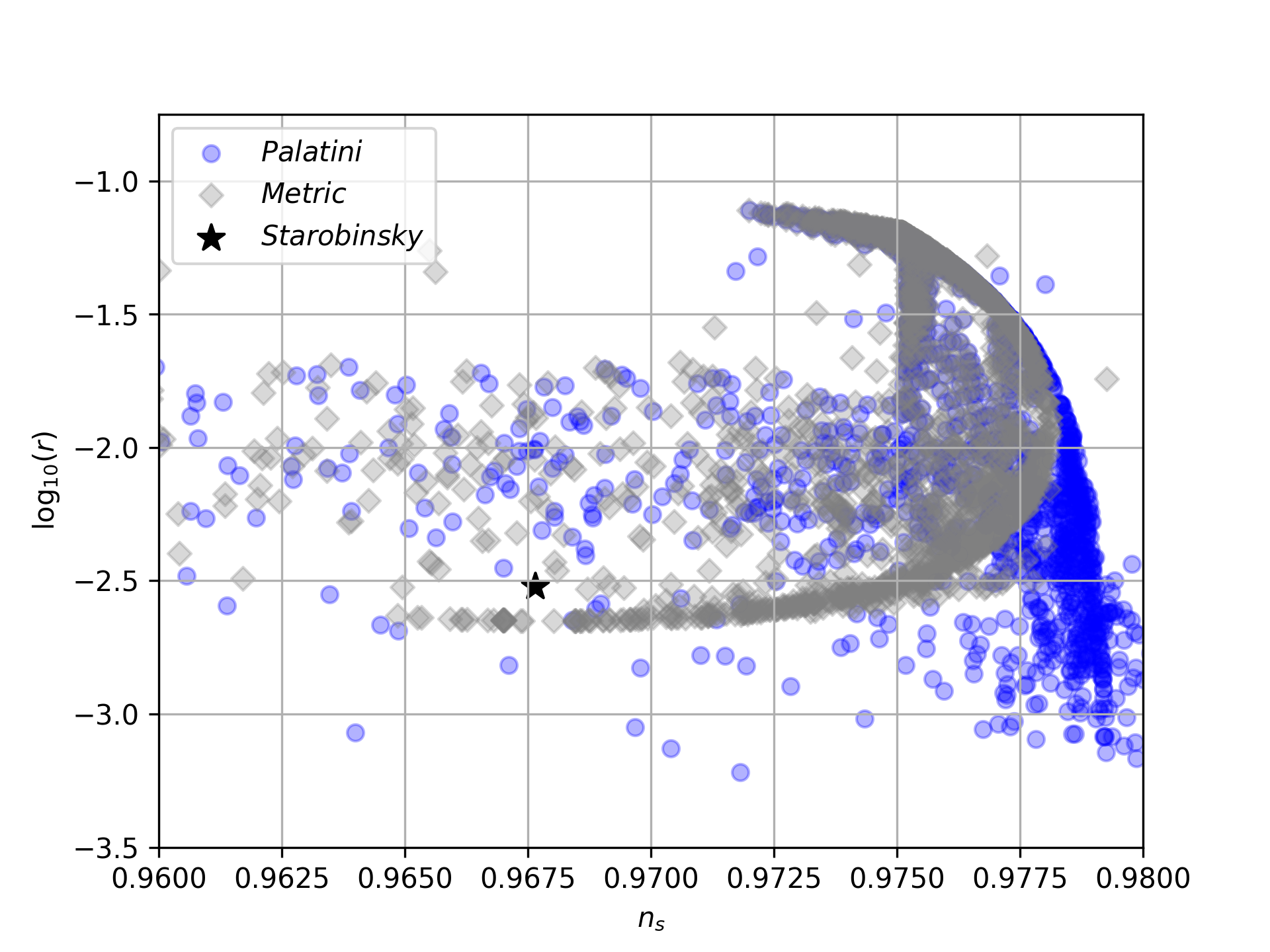}
        	\includegraphics[width=0.45\textwidth]{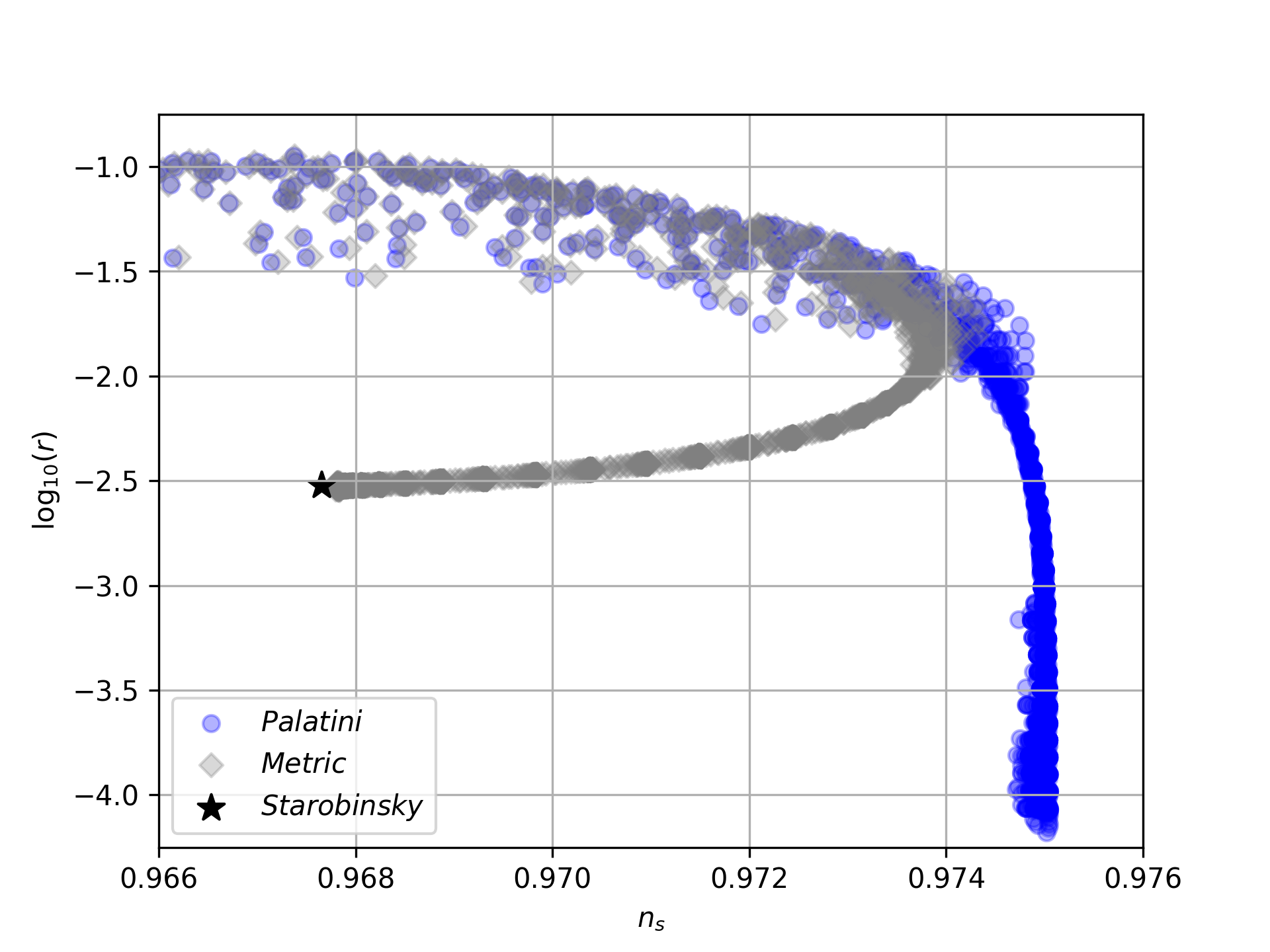}
        	\includegraphics[width=0.45\textwidth]{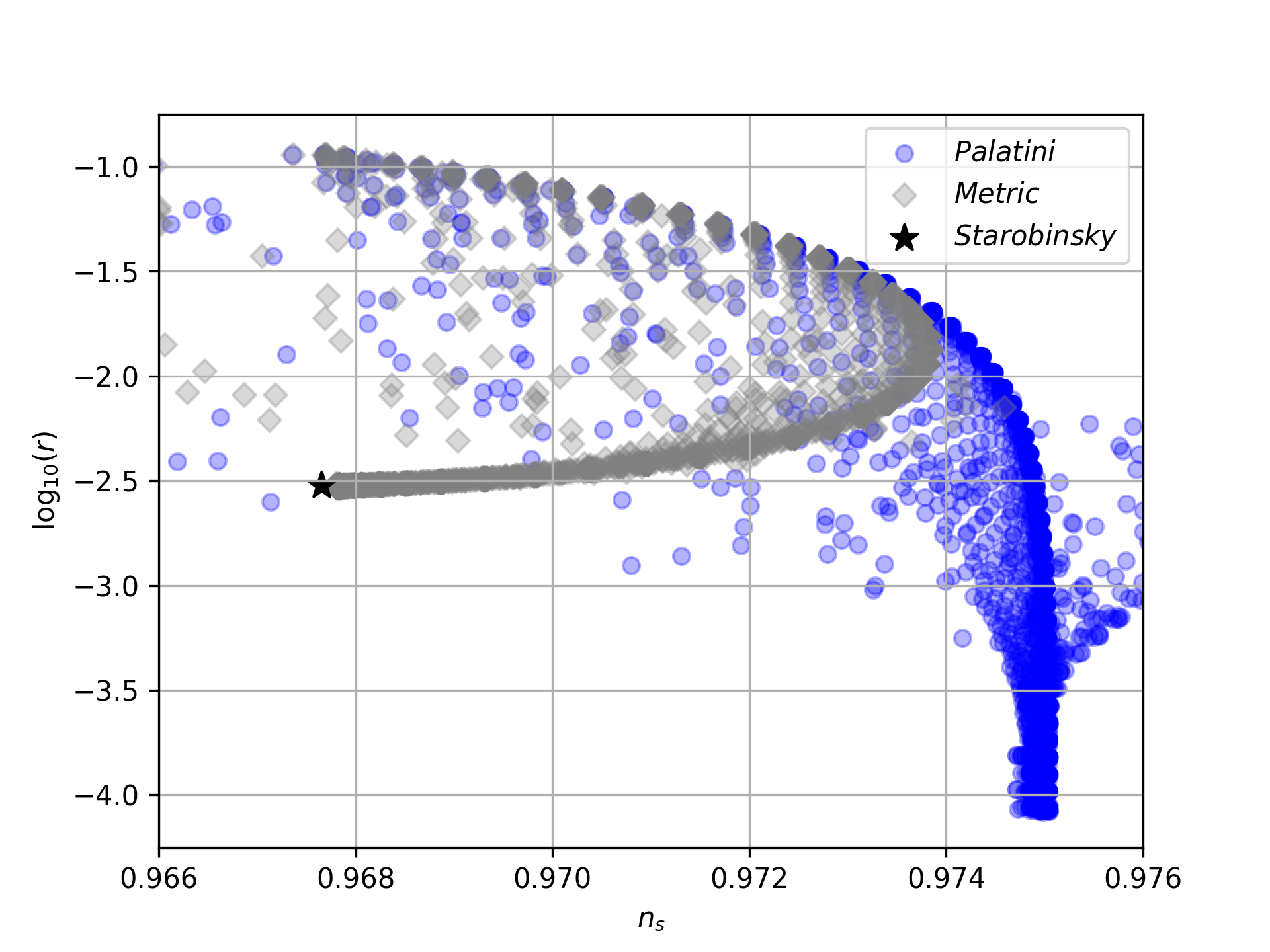}
        	\includegraphics[width=0.45\textwidth]{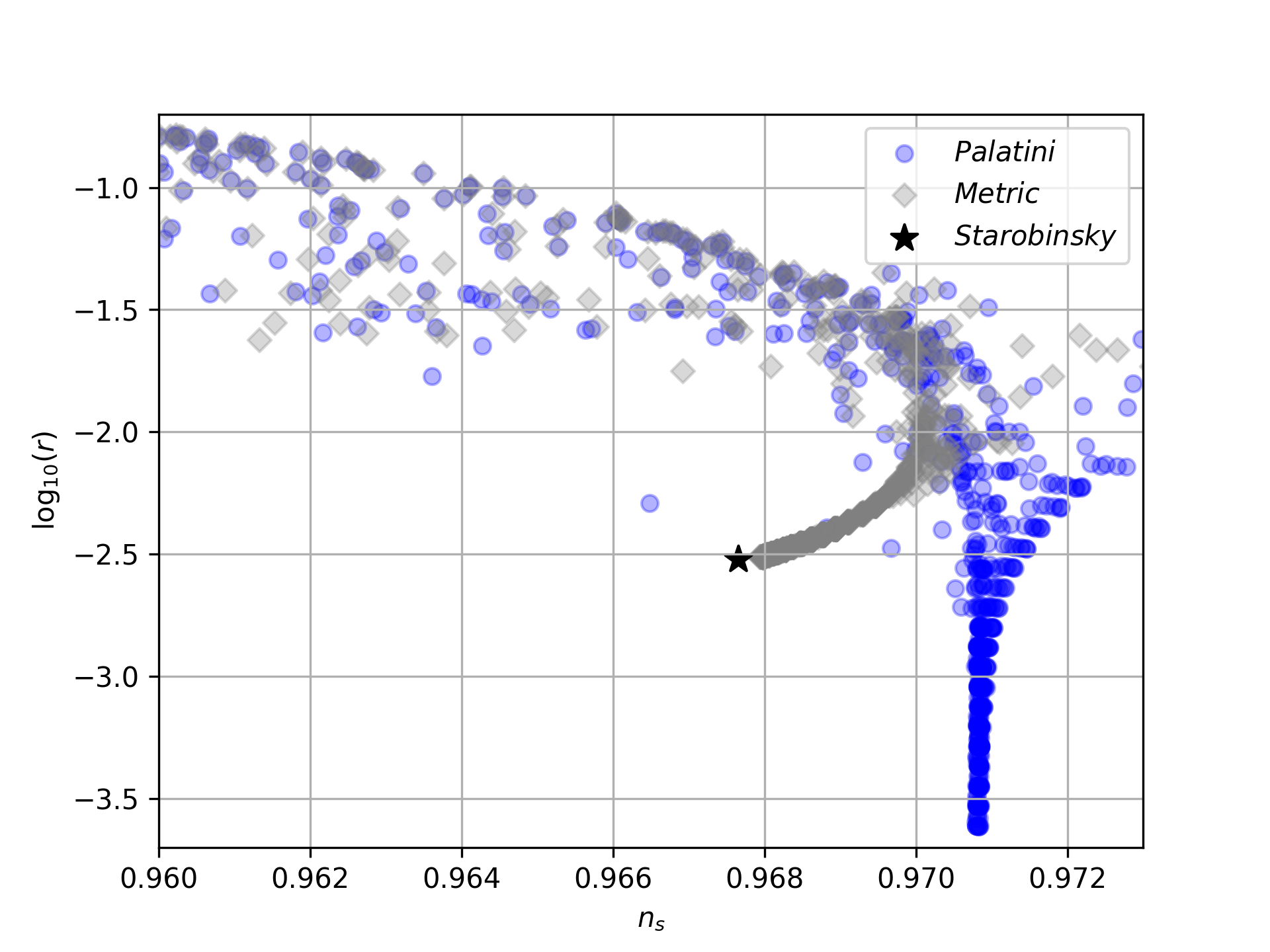}
        	\includegraphics[width=0.45\textwidth]{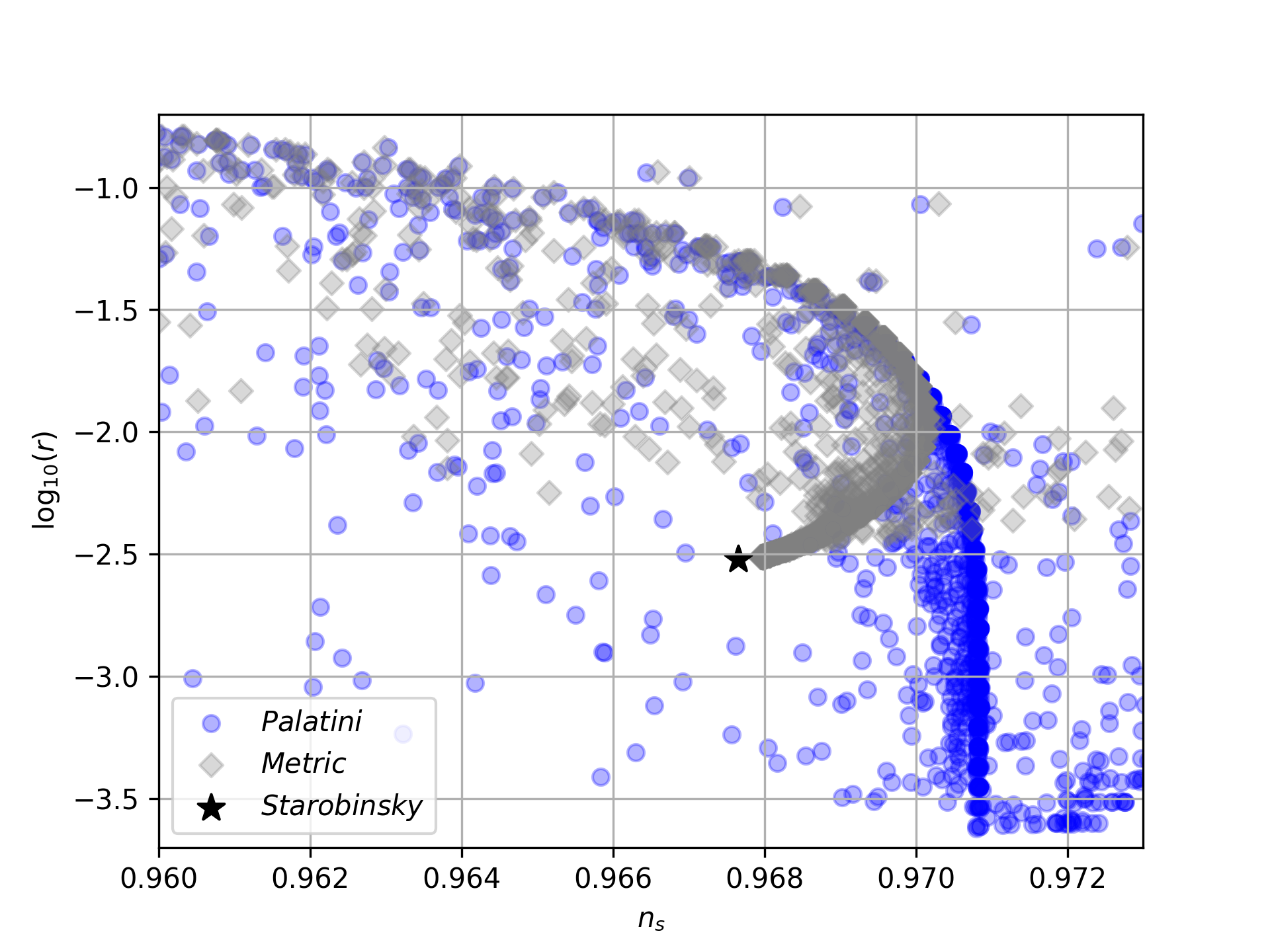}
        	\includegraphics[width=0.45\textwidth]{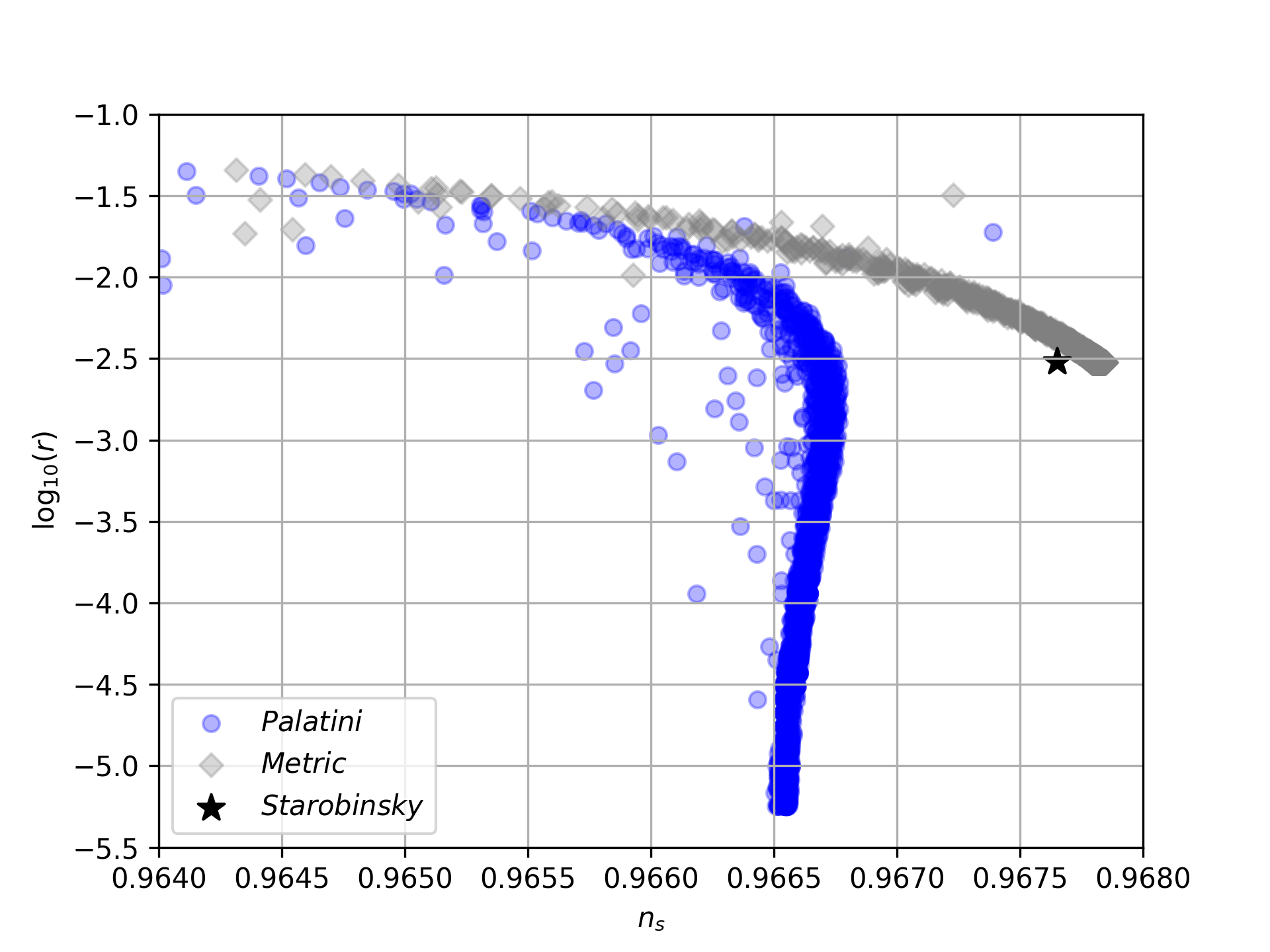}
        	\includegraphics[width=0.45\textwidth]{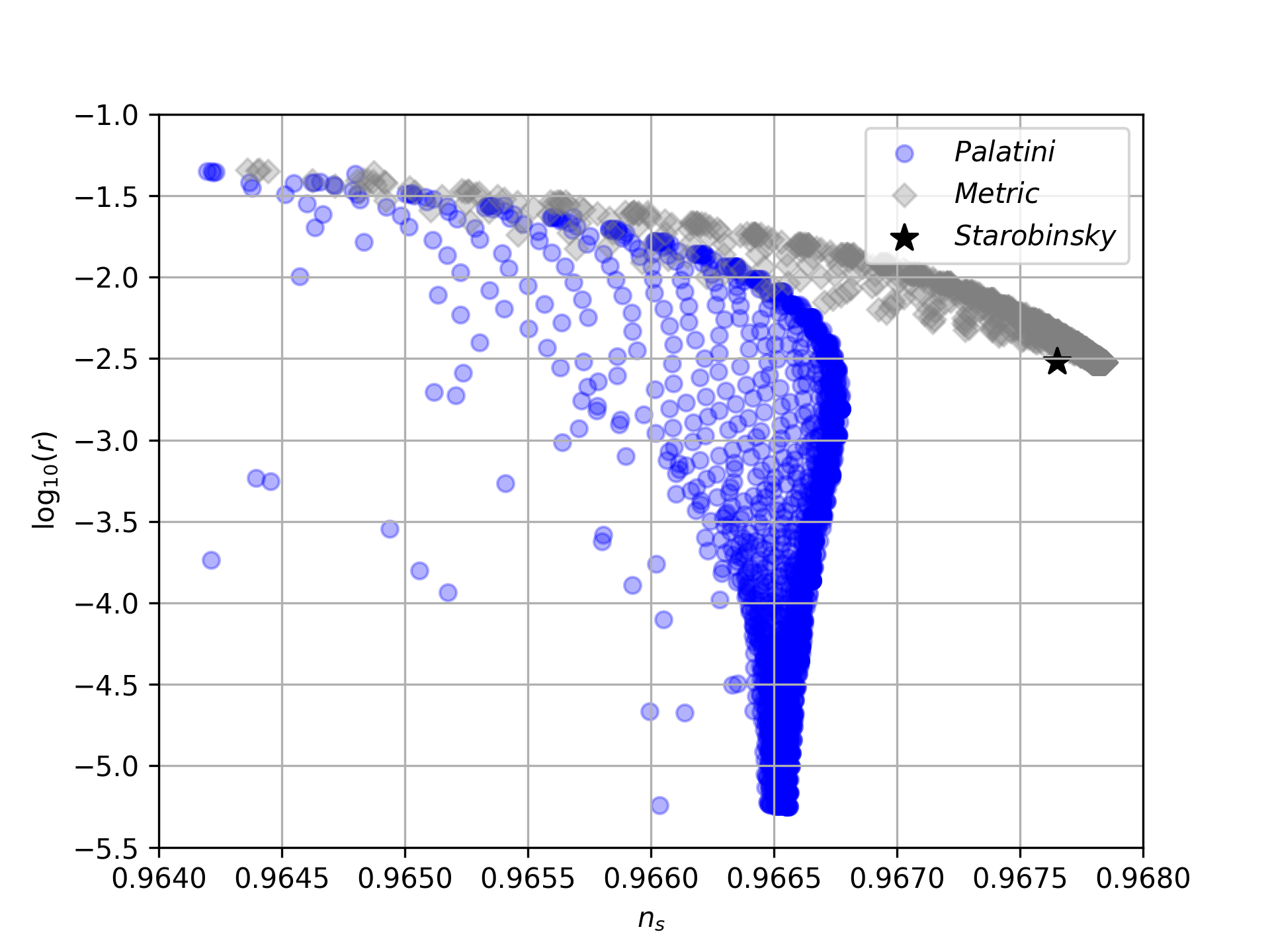}
	\caption{Predictions for $n_s$ and $r$ in metric (grey) and Palatini gravity (blue). The panels are the same as in Fig. \ref{fig:ICs}.}\label{nsr}
\end{figure}

Moving now to the observables, we study the cases for which $n=(1/2,1,3/2,2)$ in Eqs. \eqref{nmc_function} and \eqref{potential}. We show the results for $n_s$ and $r$ in Fig. \ref{nsr}. We see here a clear difference between the formulations of gravity at large values of $\xi_I$, with the results for the metric case asymptoting to those of Starobinsky inflation \cite{Starobinsky:1980te}
\bea
\label{starobinsky_nsr}
n_s^{\rm M} &\simeq 1 - \frac{2}{N} 
, \\
r^{\rm M} &\simeq \frac{12}{N^2} ,
\eea
while those for Palatini do not. The Palatini case approaches vanishing $r$ at strong coupling, asymptoting to the single field case \cite{Jarv:2017azx}
\bea
n_s^{\rm P} &\simeq 1 - \left( 1+\frac{n}{2} \right) \frac{1}{N} \,, \\
r^{\rm P} &\simeq 0 \, ,
\eea

\begin{figure}
	\centering
	        \includegraphics[width=0.45\textwidth]{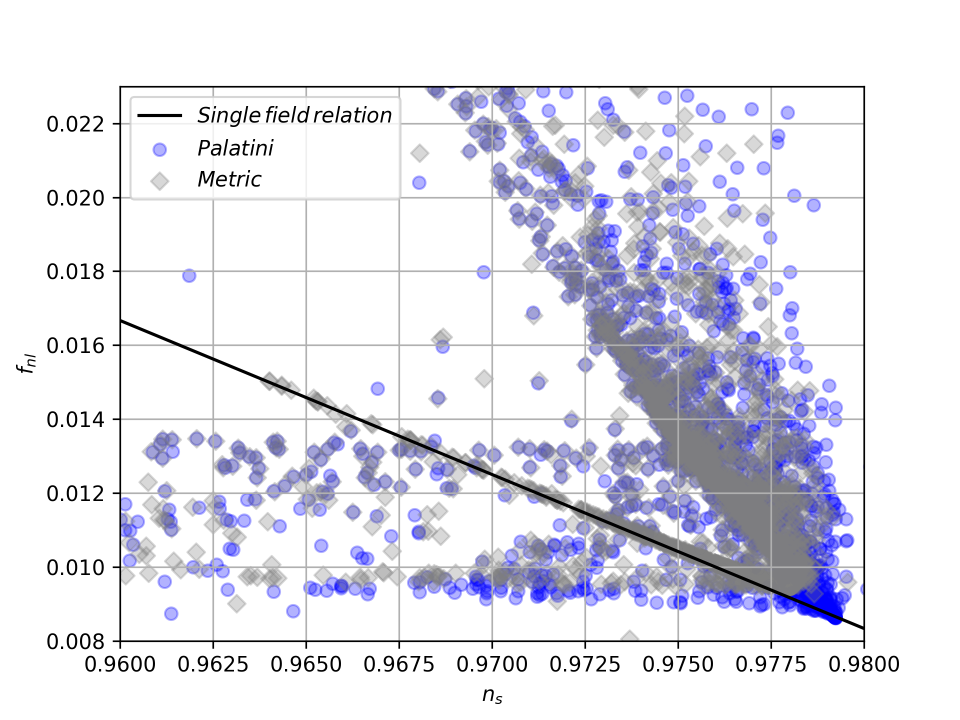}
        	\includegraphics[width=0.45\textwidth]{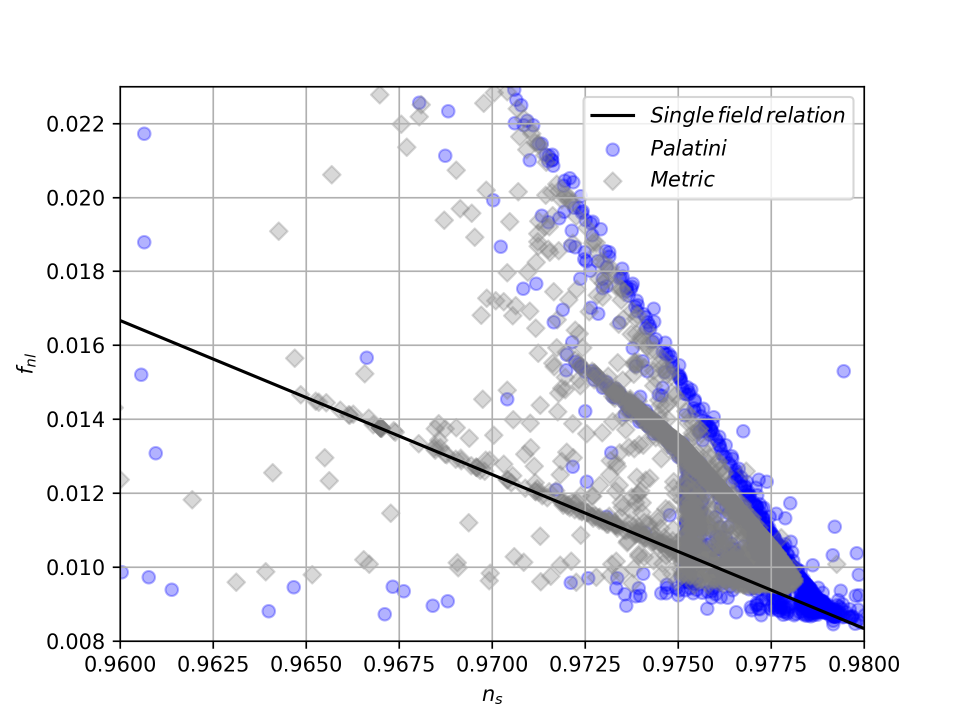}
        	\includegraphics[width=0.45\textwidth]{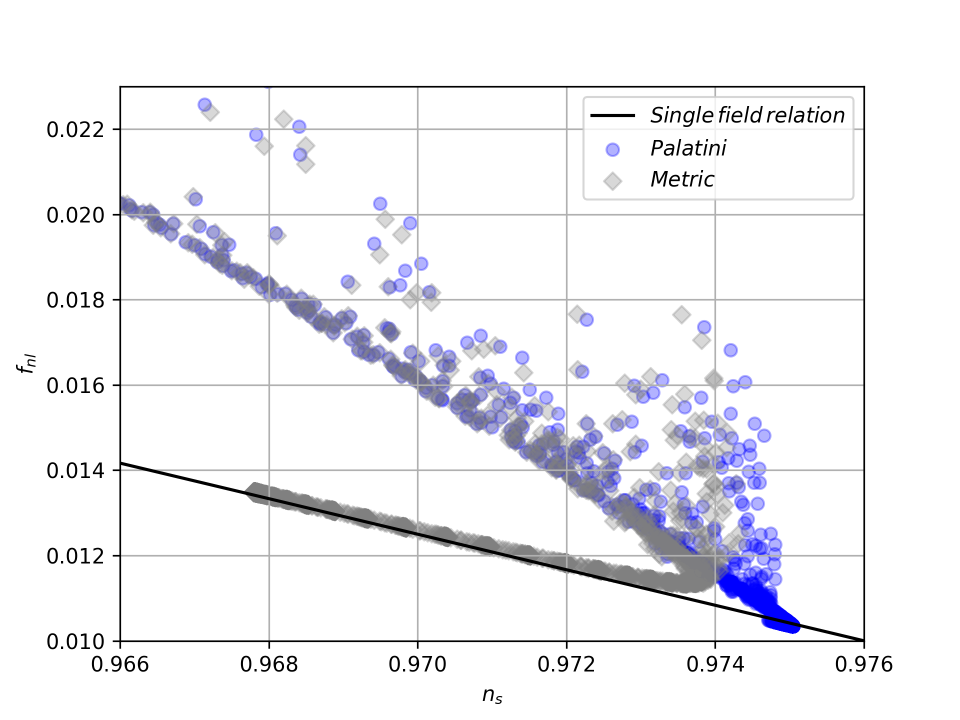}
        	\includegraphics[width=0.45\textwidth]{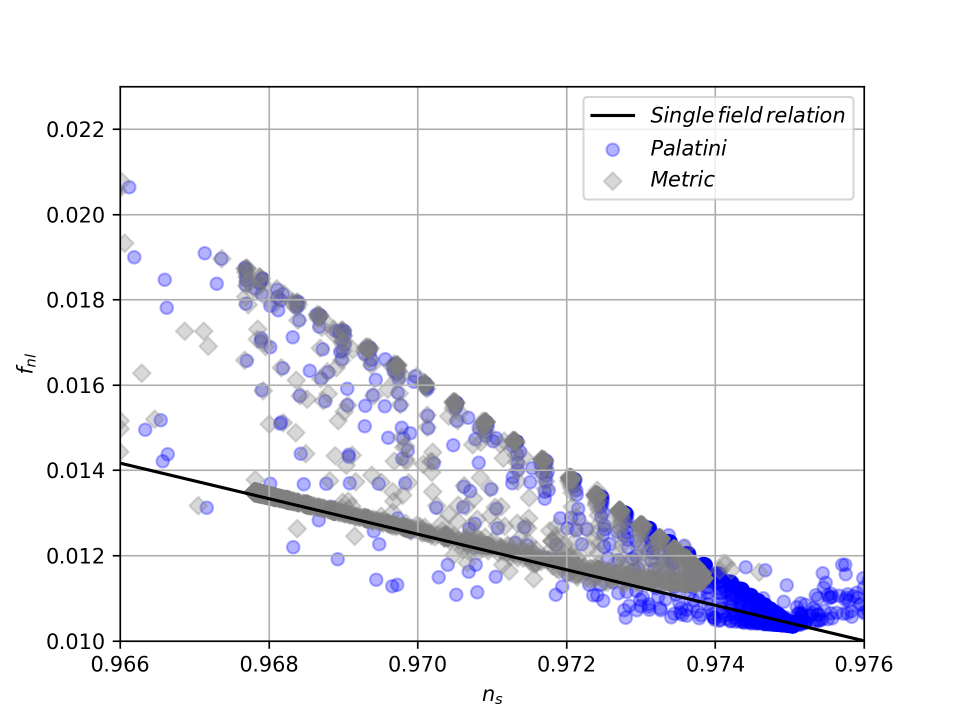}
        	\includegraphics[width=0.45\textwidth]{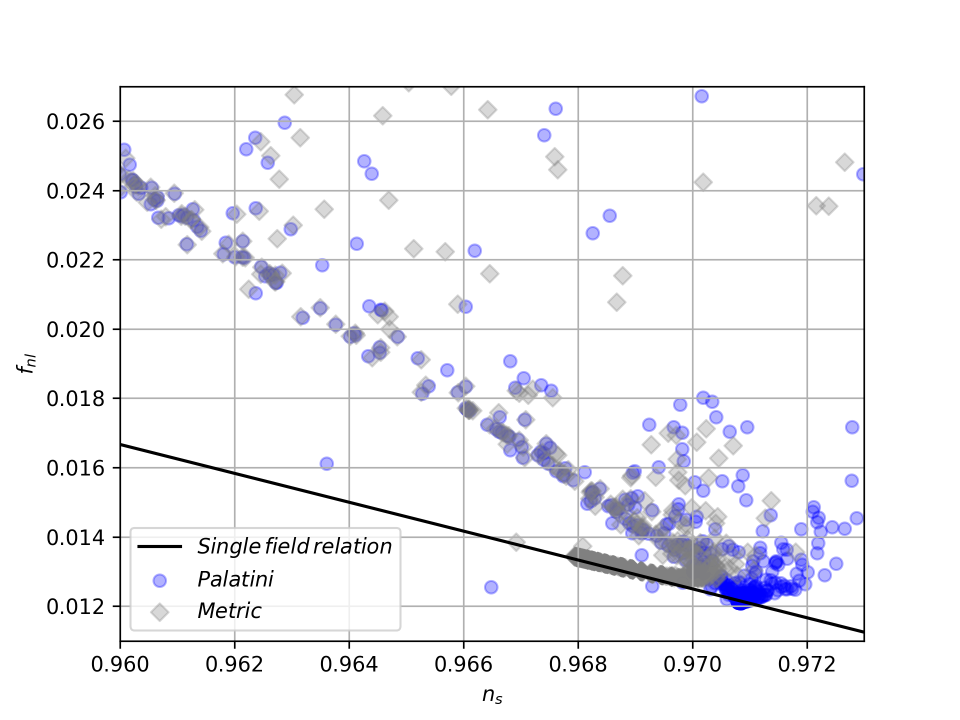}
        	\includegraphics[width=0.45\textwidth]{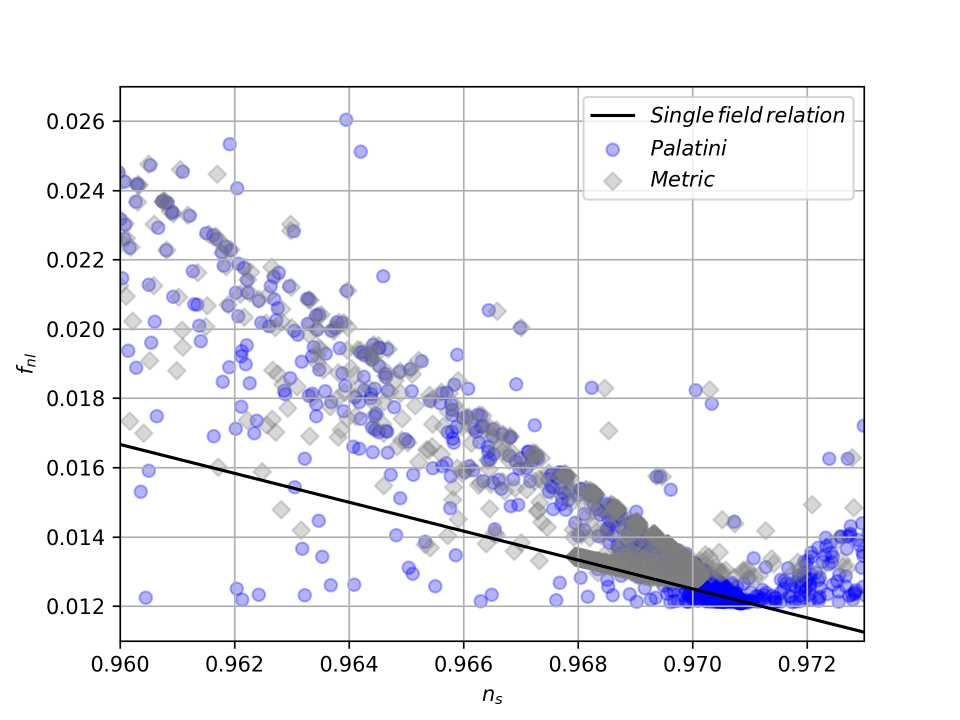}
        	\includegraphics[width=0.45\textwidth]{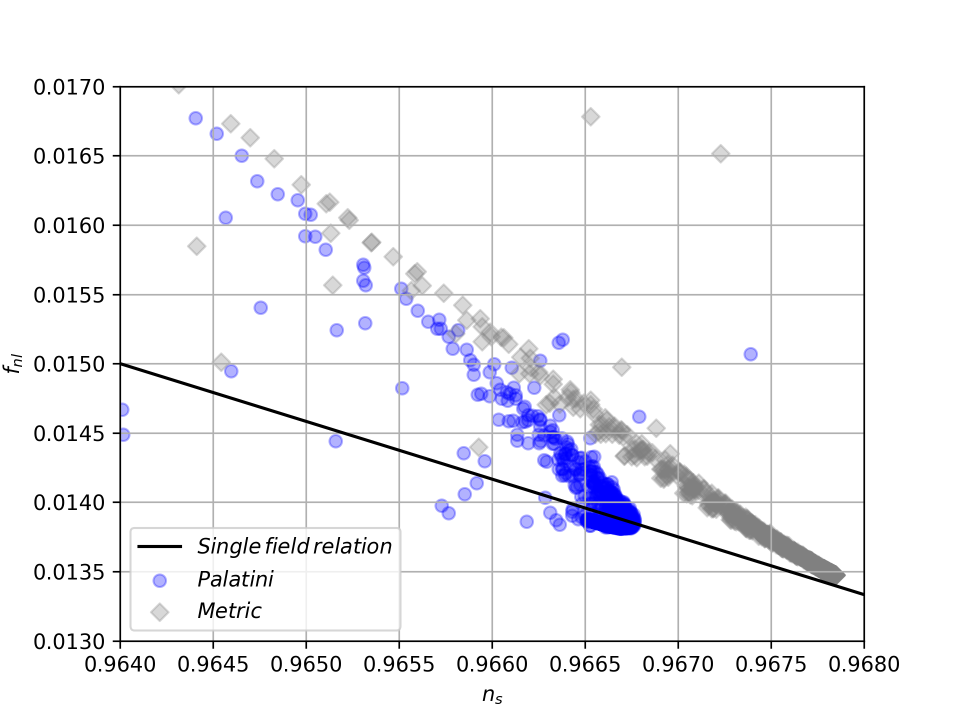}
        	\includegraphics[width=0.45\textwidth]{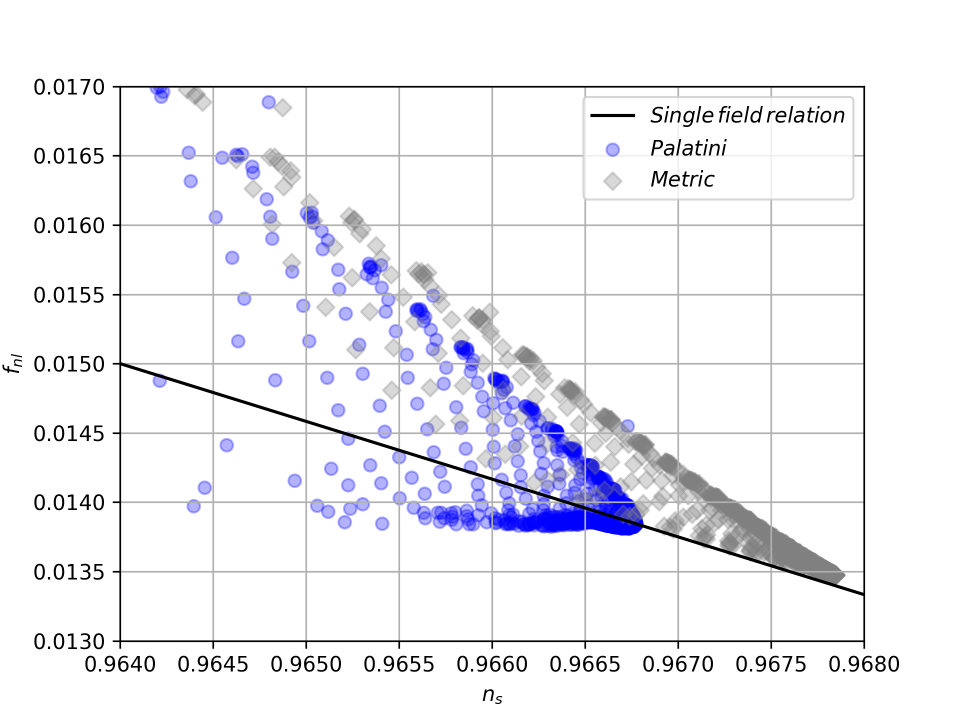}
	\caption{Predictions for $n_s$ and $f_{\rm NL}$ in metric (grey) and Palatini gravity (blue). The panels are the same as in Fig. \ref{fig:ICs}.}\label{nsfnl}
\end{figure}

\begin{figure}
	\centering
	        \includegraphics[width=0.45\textwidth]{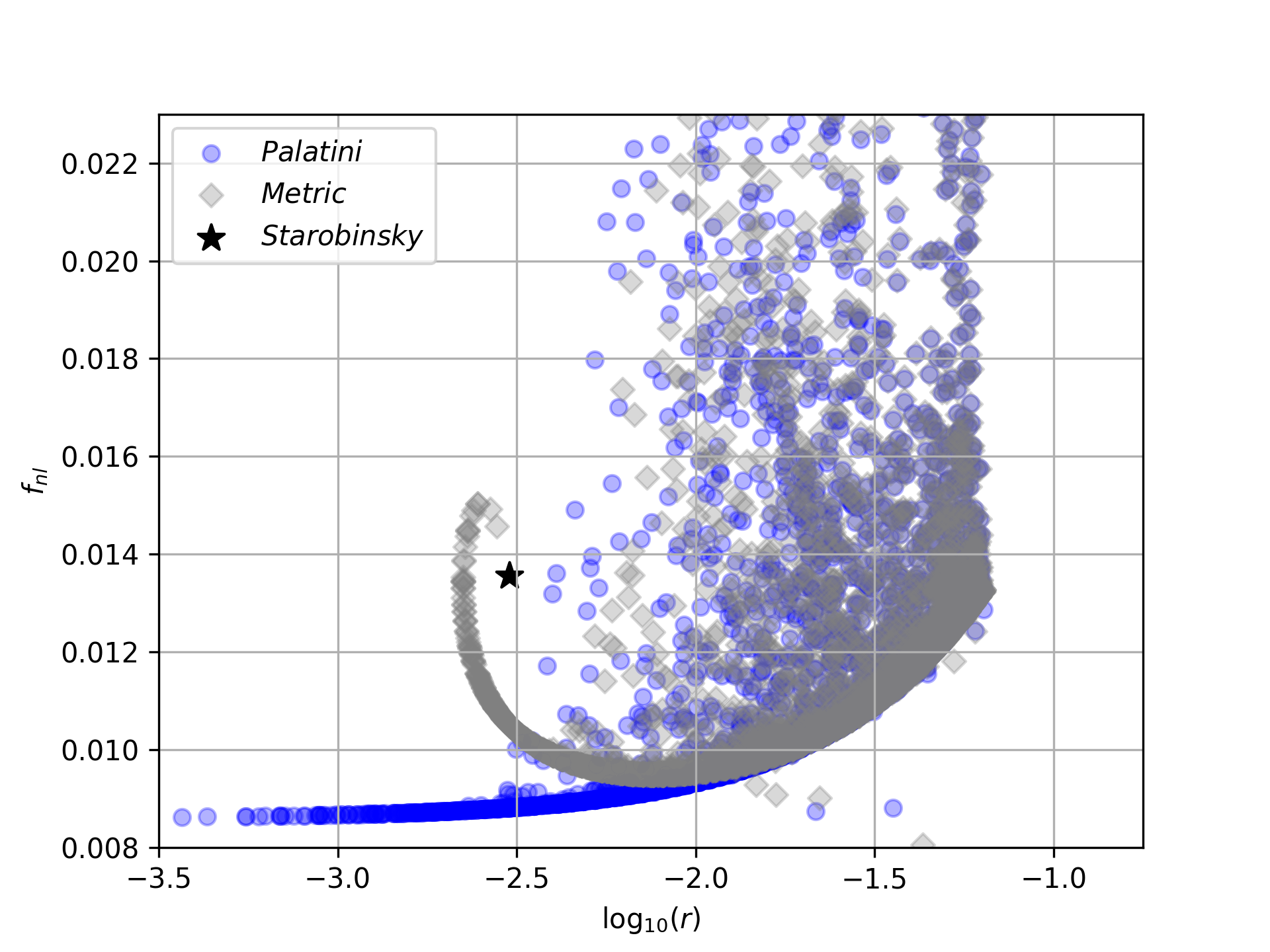}
        	\includegraphics[width=0.45\textwidth]{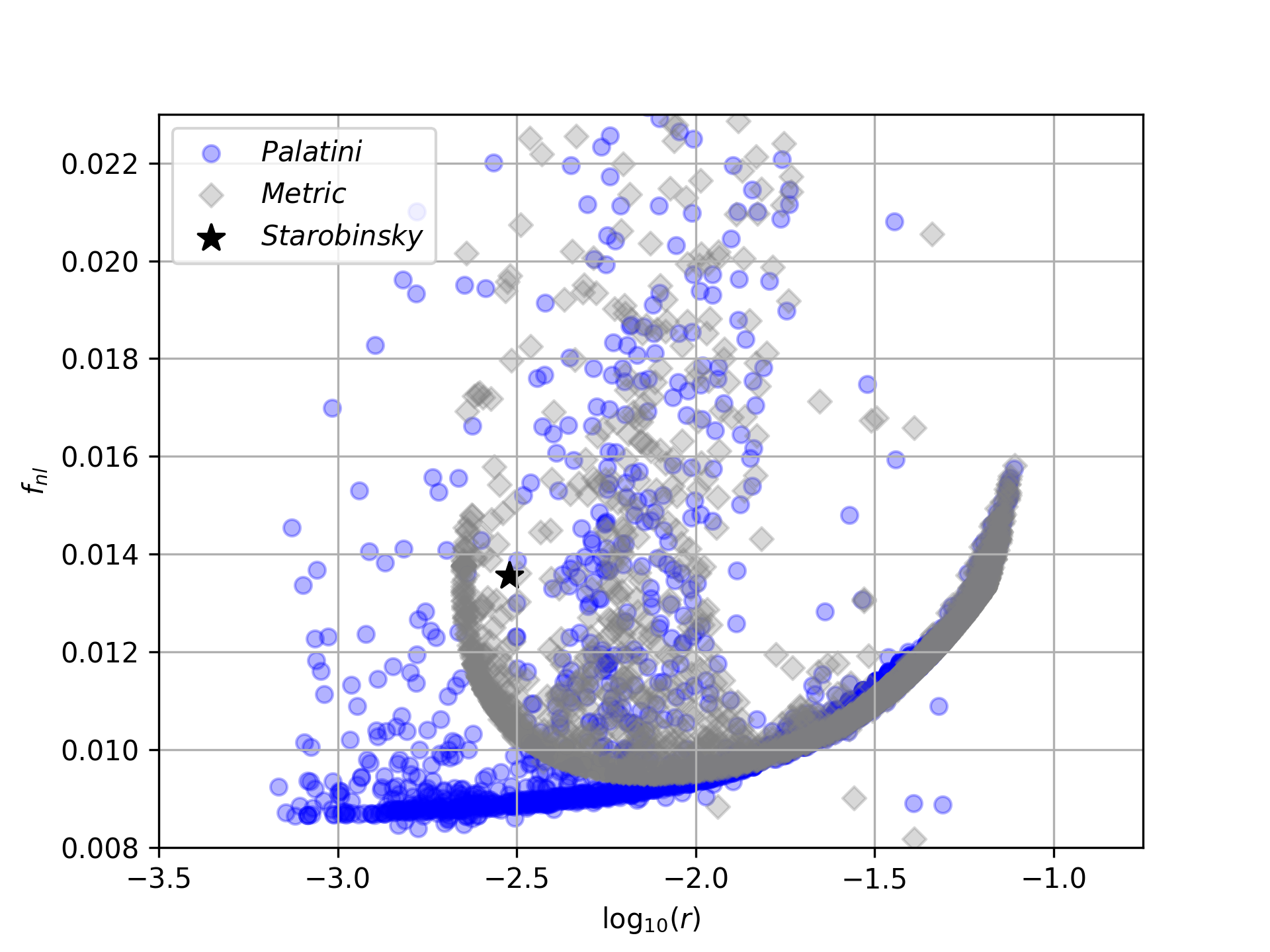}
        	\includegraphics[width=0.45\textwidth]{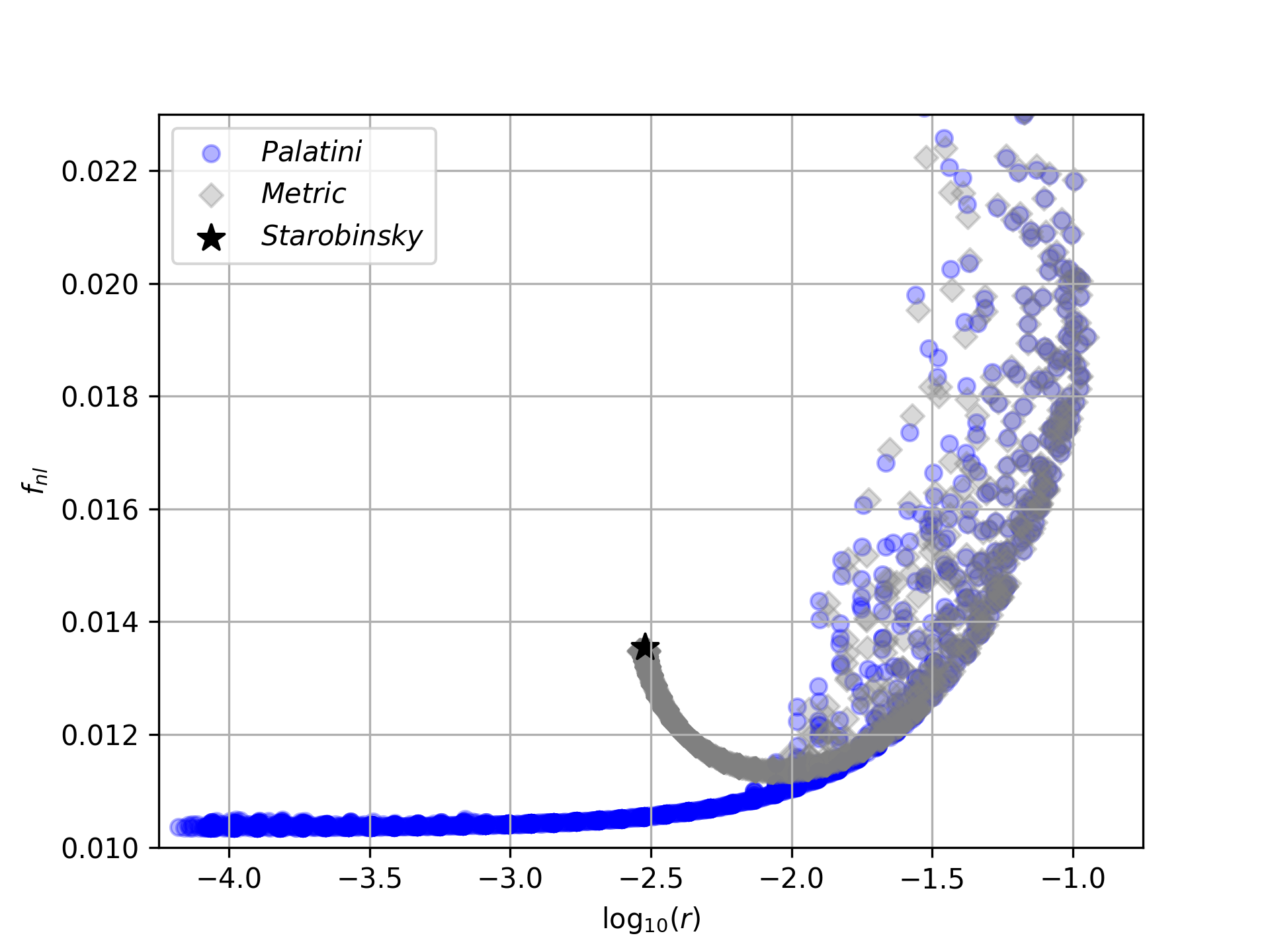}
        	\includegraphics[width=0.45\textwidth]{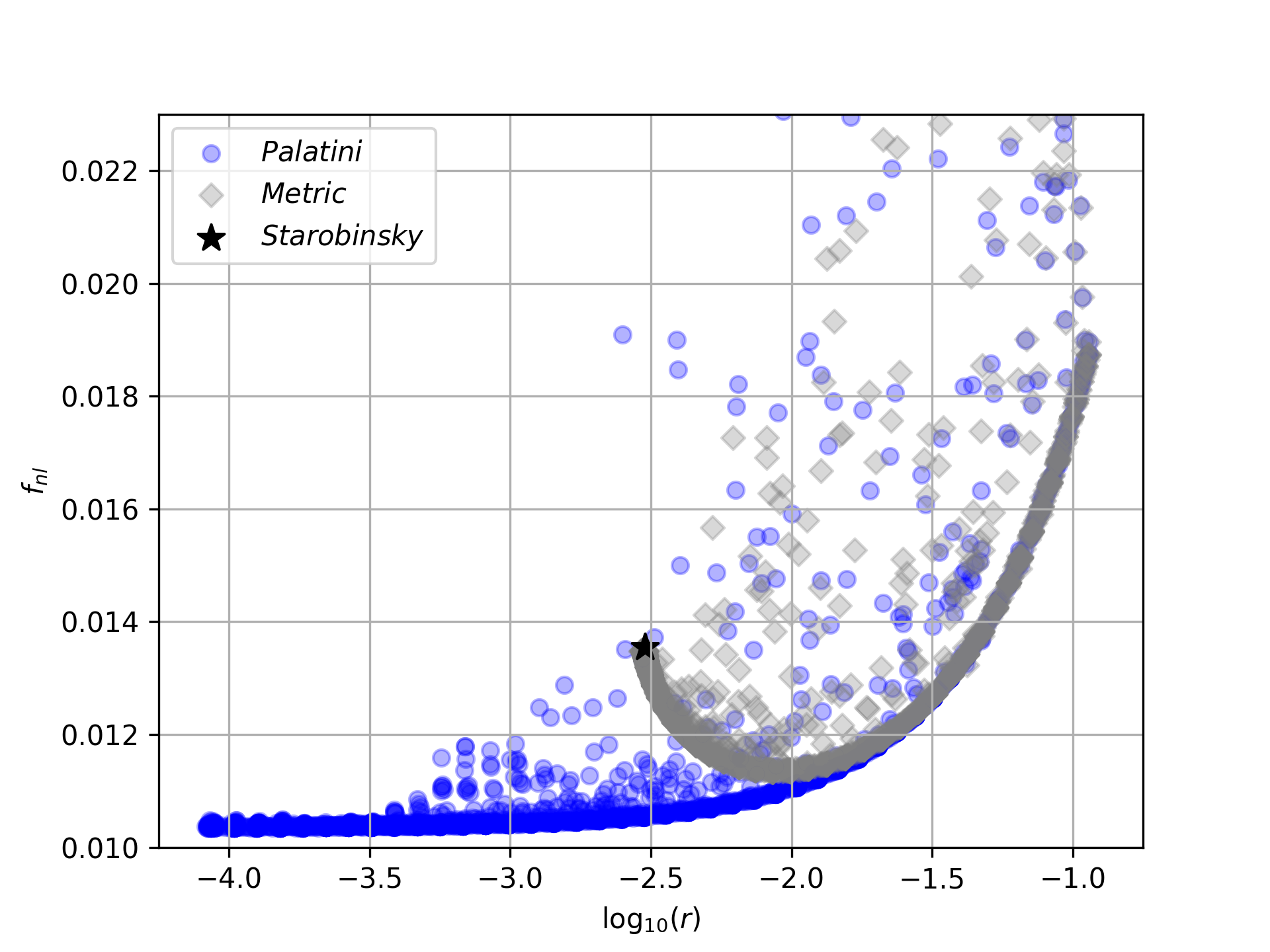}
        	\includegraphics[width=0.45\textwidth]{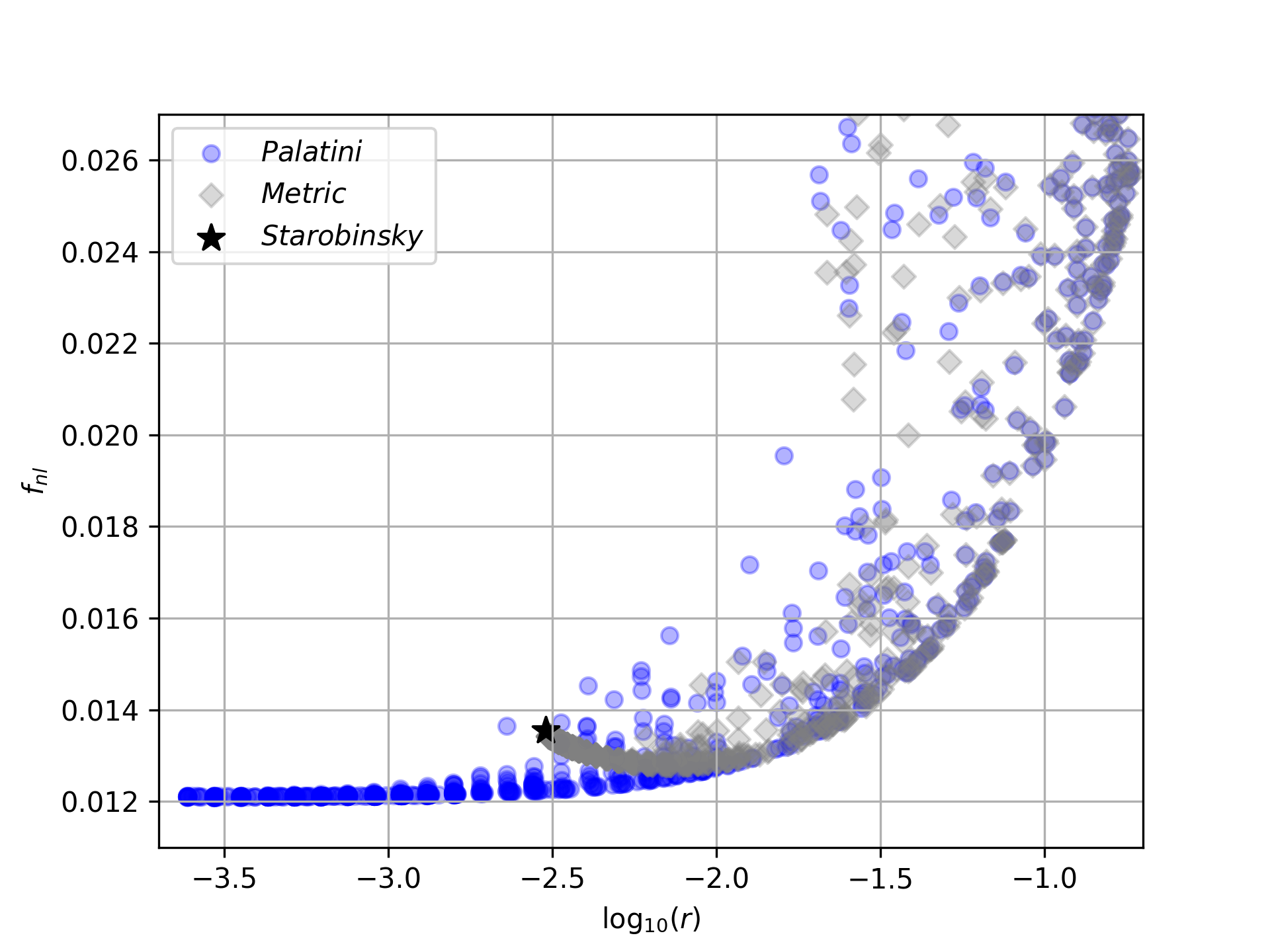}
        	\includegraphics[width=0.45\textwidth]{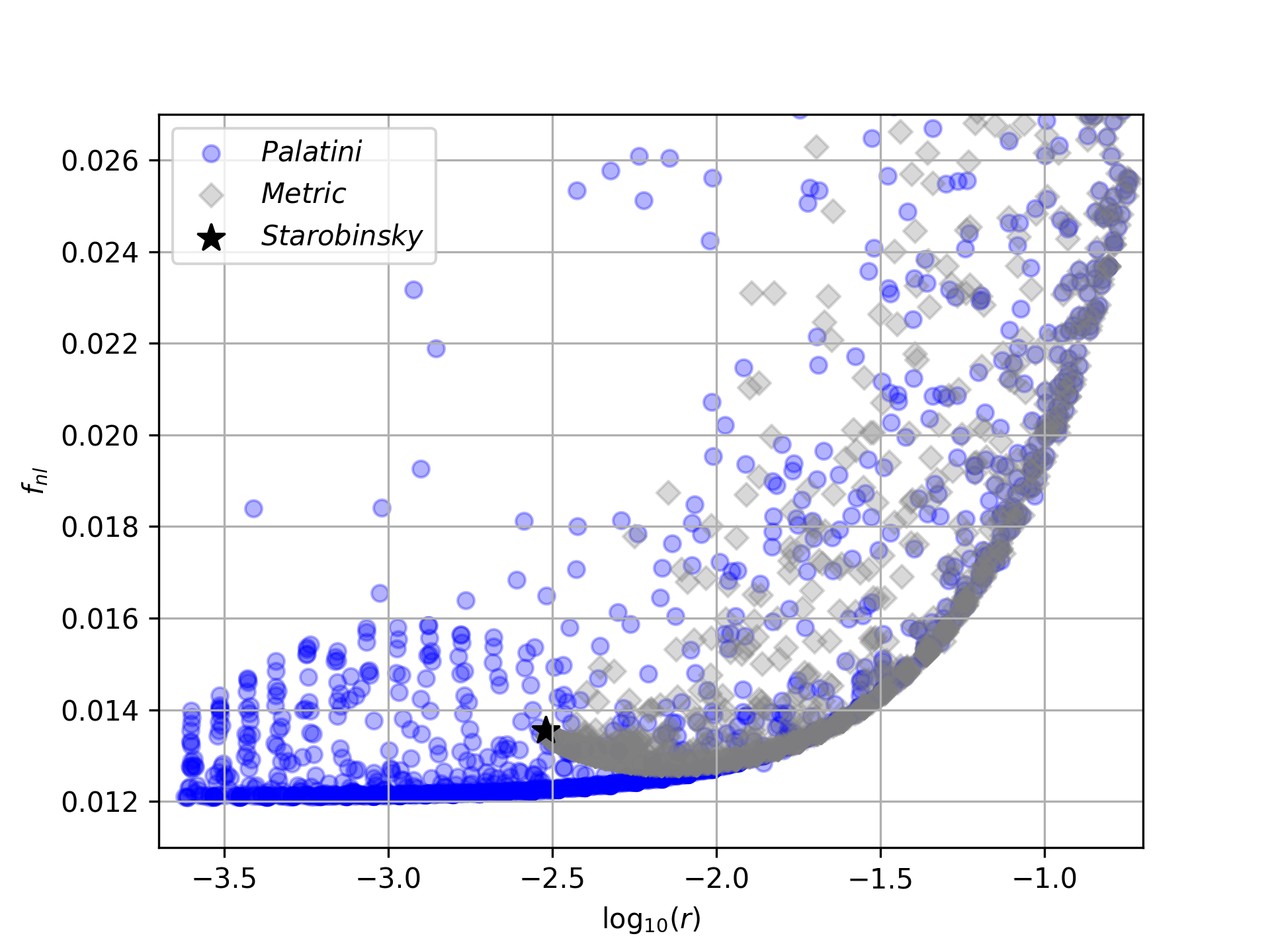}
        	\includegraphics[width=0.45\textwidth]{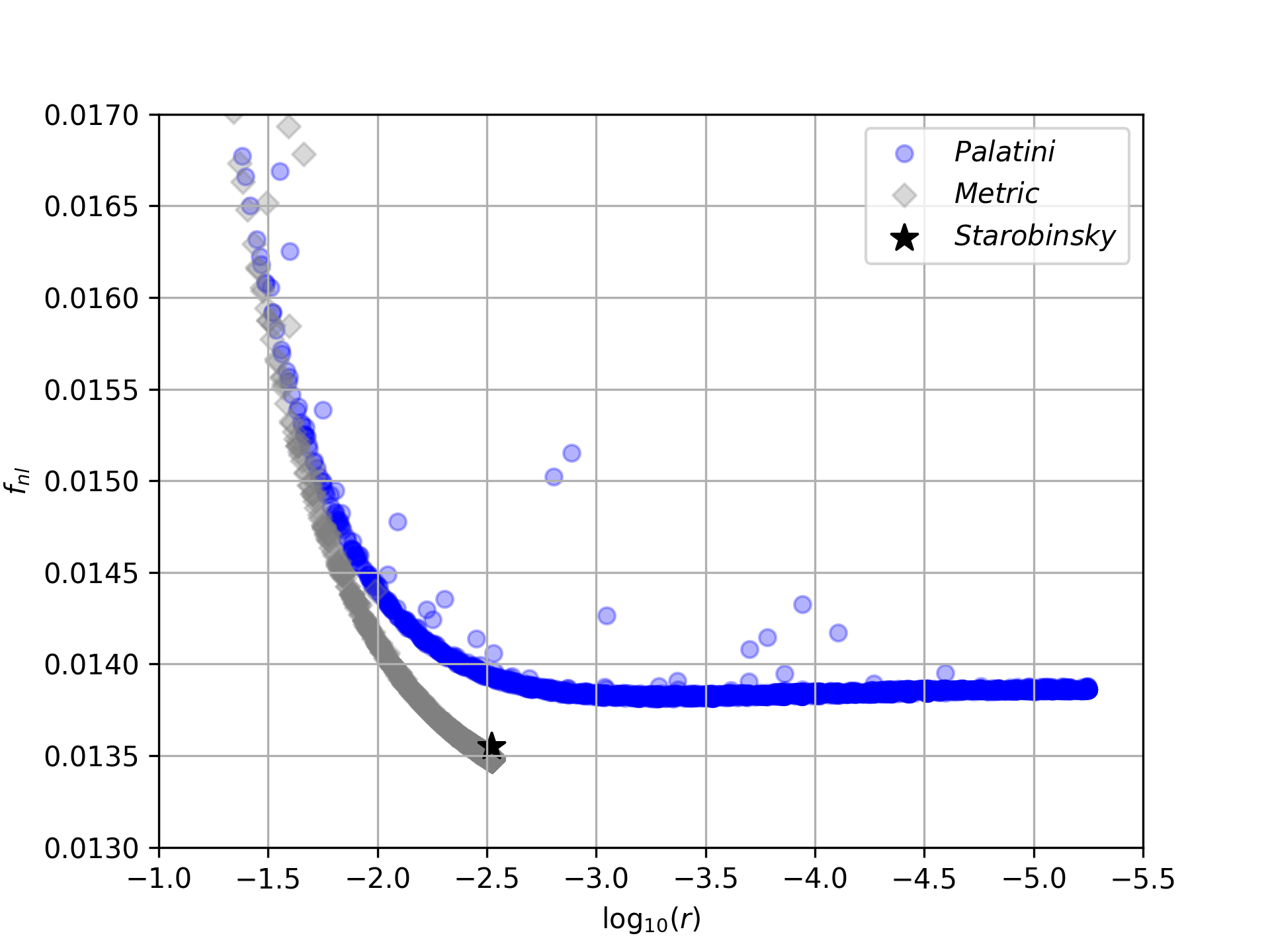}
        	\includegraphics[width=0.45\textwidth]{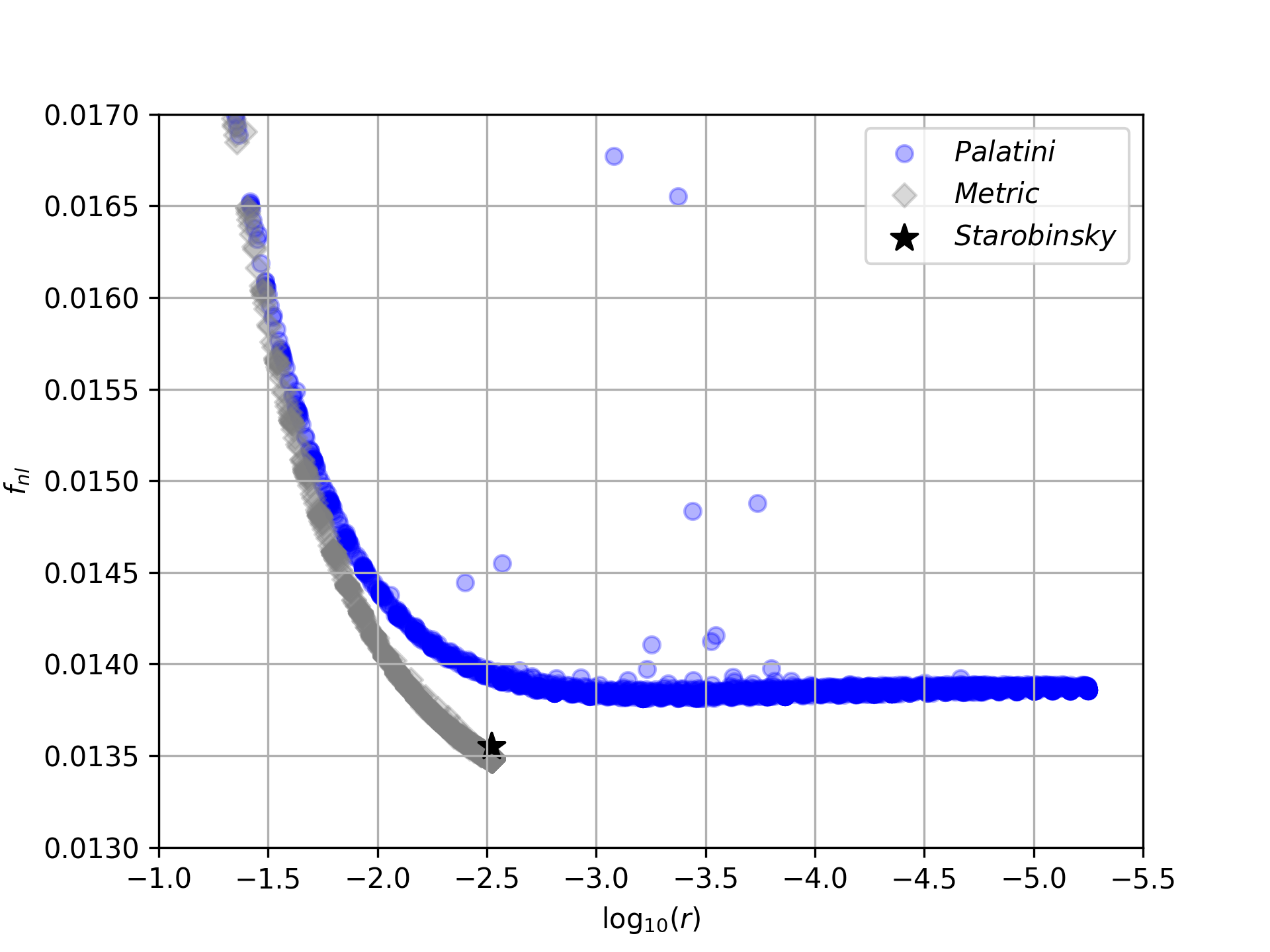}
	\caption{Predictions for $r$ and $f_{\rm NL}$ in metric (grey) and Palatini gravity (blue). The panels are the same as in Fig. \ref{fig:ICs}.}\label{rfnl}
\end{figure}

However, we find that in the Palatini case the results converge to a non-zero value of $f_{\rm NL}$, which is different from that of the metric case. The results are shown in Figs. \ref{nsfnl} and \ref{rfnl} along with lines corresponding to the Maldacena's consistency relation $f_{\rm NL}=5/12(1-n_s)$ \cite{Maldacena:2002vr} for the single-field case\footnote{One expects Maldacena's relation to hold for squeezed configurations of the reduced bispectrum, while here we are plotting the reduced bispectrum in the equilateral limit. However, in canonical single field models in which $\epsilon \ll \eta$, which is the case for the single field limit here, the bispectrum is very close to local and the reduced bispectrum is almost the same in all configurations. This is why our plot for $f_{\rm NL}$ against $n_s$ follows so closely the Maldacena relation.}. We see that the values of $f_{\rm NL}$ converge to the single-field result at strong coupling for both Palatini and metric gravity, confirming the general trend that the multifield results mimic those of the single-field case in the strong coupling limit.

We see that all multifield models considered in the present paper reduce to an effective single-field model at the limit of strong coupling. In the metric case this generalizes the earlier findings in the literature\footnote{Outside the context of inflation, similar single field behaviour has been found in other scenarios with non-minimally coupled multifield models~\cite{Hohmann:2016yfd}.}, as discussed in Sec. \ref{introduction}, whereas in the Palatini case the results are entirely new. We elaborate on the reason for this behaviour in the next subsection. However, we stress that the scenario was not constructed to obtain an $\alpha$-attractor model but it emerges naturally from the Jordan frame action \eqref{nonminimal_action}, which is our starting point. 

Also, note that if one considered a scenario in which the Jordan frame action included non-canonical kinetic terms of a specific kind, one would get the same result as in the present case 
where the kinetic terms are canonical in the Jordan frame but where the conformal transformation and the resulting kinetic terms in the Einstein frame depend upon the assumed gravitational degrees of freedom. 
For example, the models that we consider in the Palatini formalism
are equivalent to 
non-canonical scalar-tensor theories in the metric formalism~\cite{Sotiriou:2006hs}. However, as discussed in Sec. \ref{introduction}, non-minimal couplings to gravity should be seen not as an {\it ad-hoc} addition to inflationary models but as a generic ingredient of coherent model frameworks, generated by quantum corrections in a curved space-time. It is by this notion that one can say that the differences observed between the cases which we call `metric' and `Palatini' are indeed in the underlying theory of gravity, i.e. whether the space-time connection was determined by the metric only, or both the metric and the inflaton field(s). Our study therefore reveals an interesting subtlety in a broad class of models where the scalar potential is multidimensional and the fields are non-minimally coupled to gravity.

Alternatively, one can view this work as a more detailed way to answer the question `What are the predictions of a given model of inflation?'. As shown in the present paper, they clearly depend on the choice of the gravitational degrees of freedom, even though usually such a choice is not considered to be part of models of inflation. It is therefore important to investigate all possibilities concerning the physics at high energies, as one cannot distinguish between the metric and Palatini formalisms at late times. Detailed studies of non-minimally coupled models are therefore interesting not only from the inflationary point of view, but also because they may provide for a way to distinguish between different formulations of gravity. 

\subsection{Multifield effects}
\label{fs_curvature}

Having discussed the general trends in the previous sections, we now discuss some of the effects of having multiple fields. The first effect we study is the dependence on the hierarchy between the values for the non-minimal couplings. In order to do that, we use the polar coordinates in parameter space introduced in Eq.~\eqref{polarxi} and test the evolution of the observables depending on $\theta$.

\begin{figure}
	\centering
        	\includegraphics[width=0.45\textwidth]{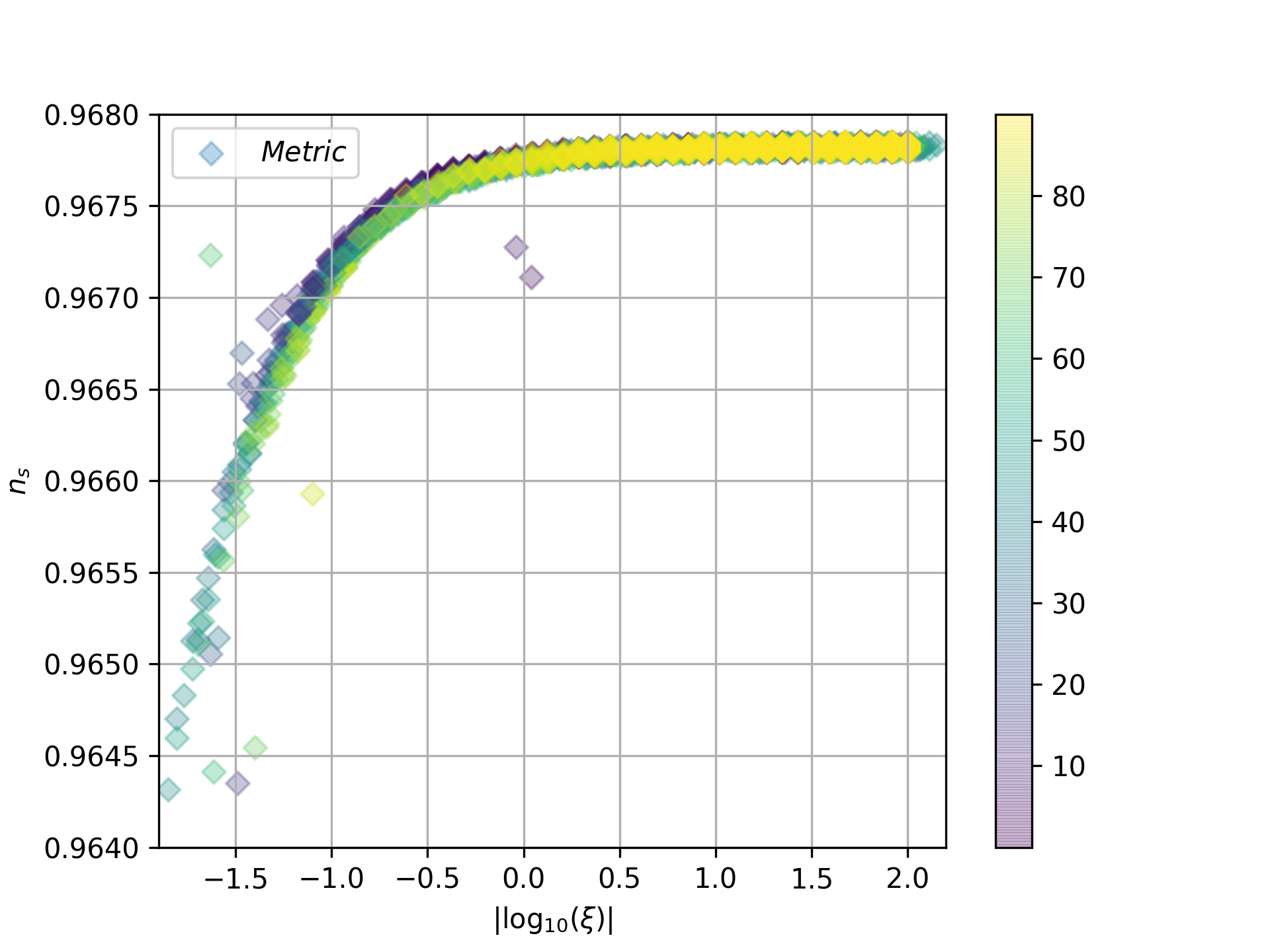}
        	\includegraphics[width=0.45\textwidth]{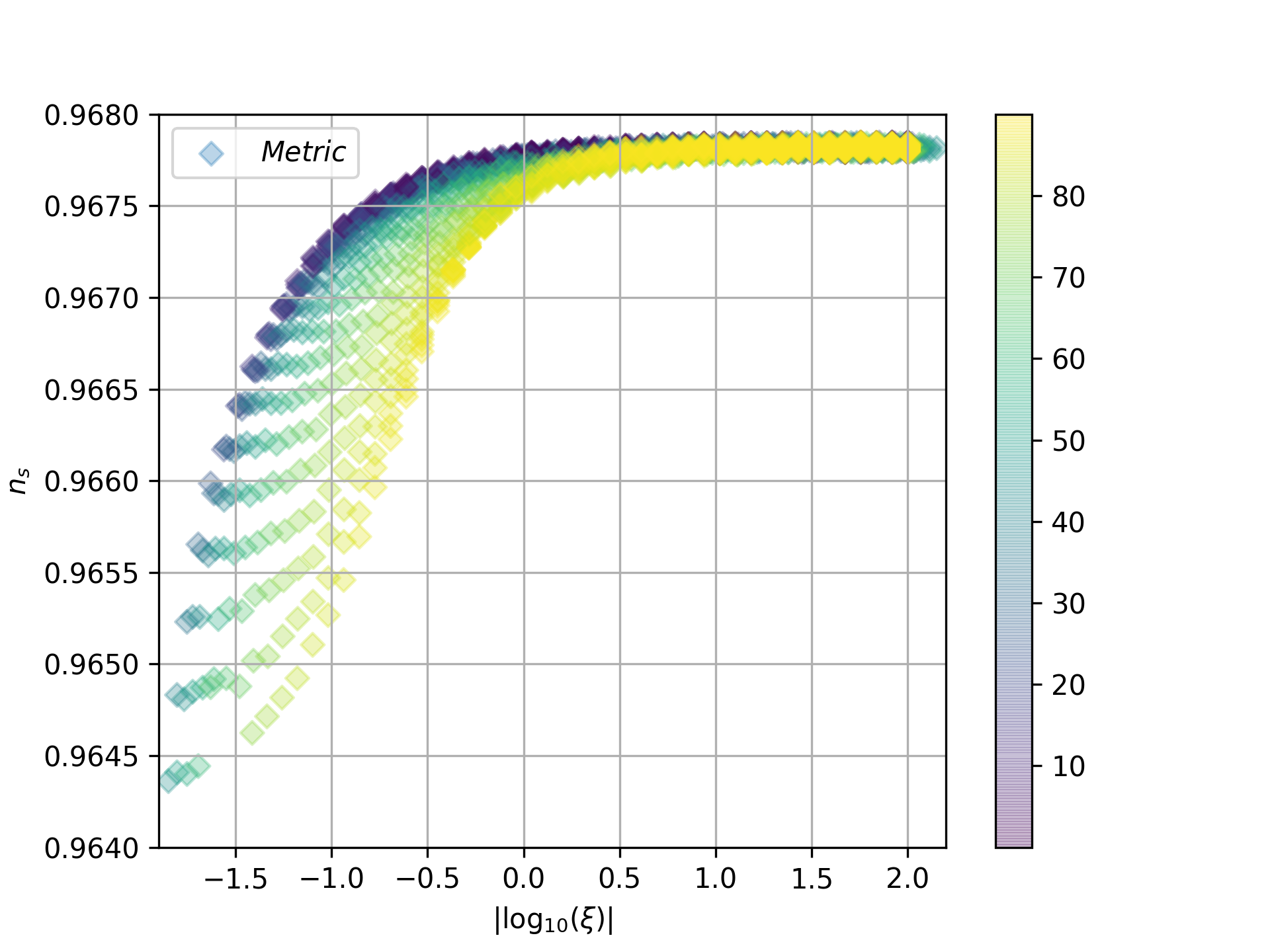}
        	\includegraphics[width=0.45\textwidth]{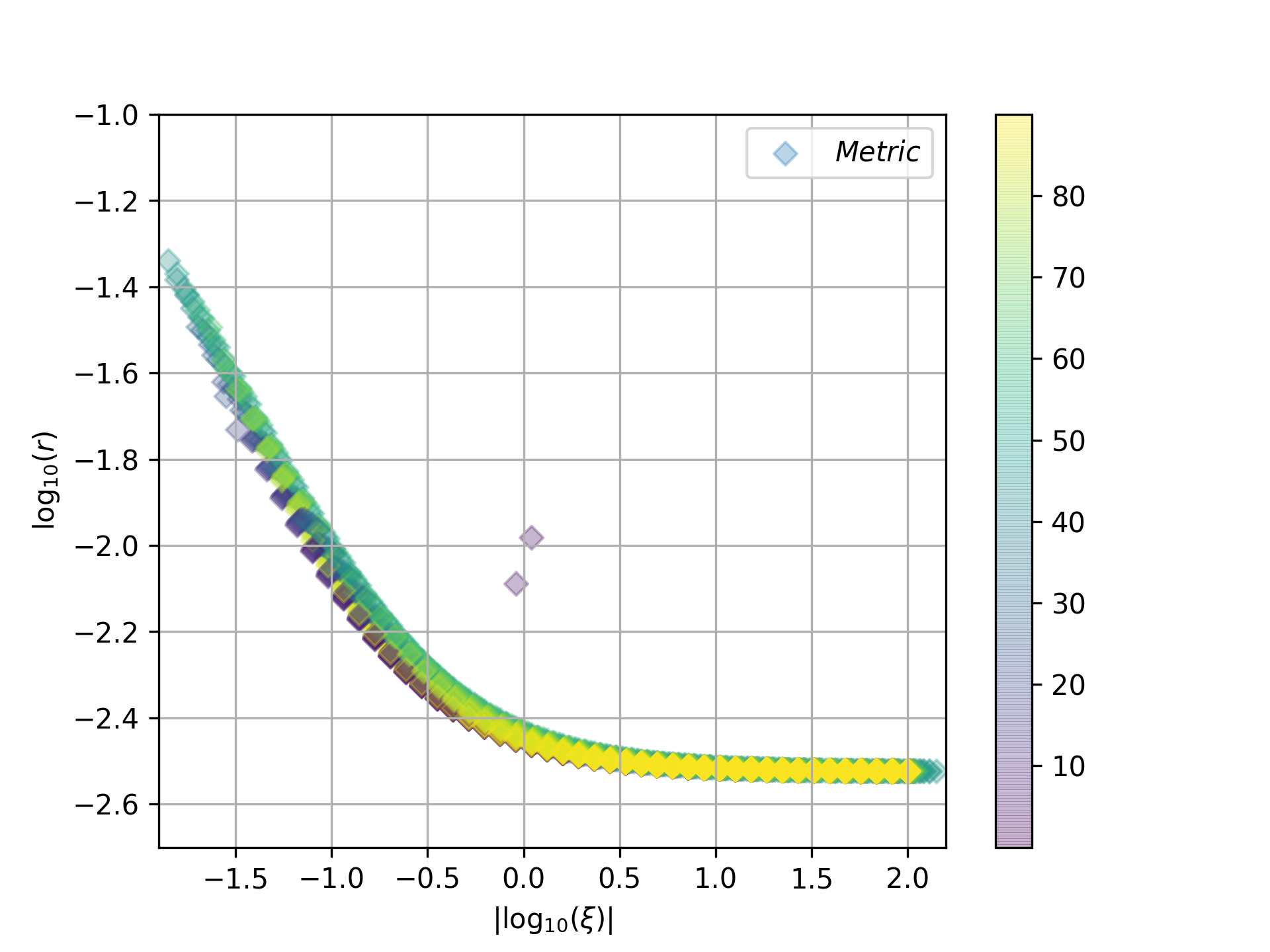}
        	\includegraphics[width=0.45\textwidth]{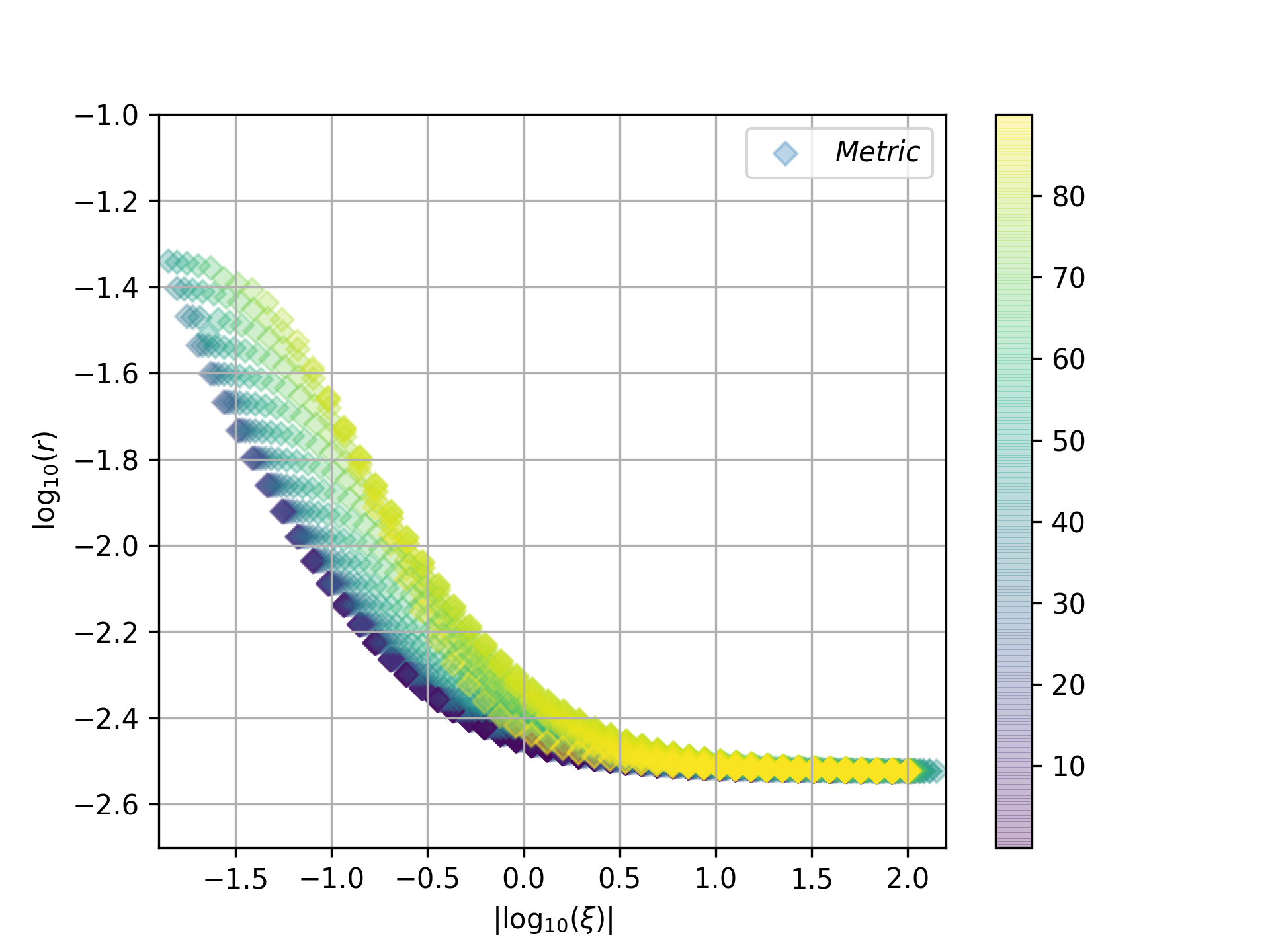}
		\includegraphics[width=0.45\textwidth]{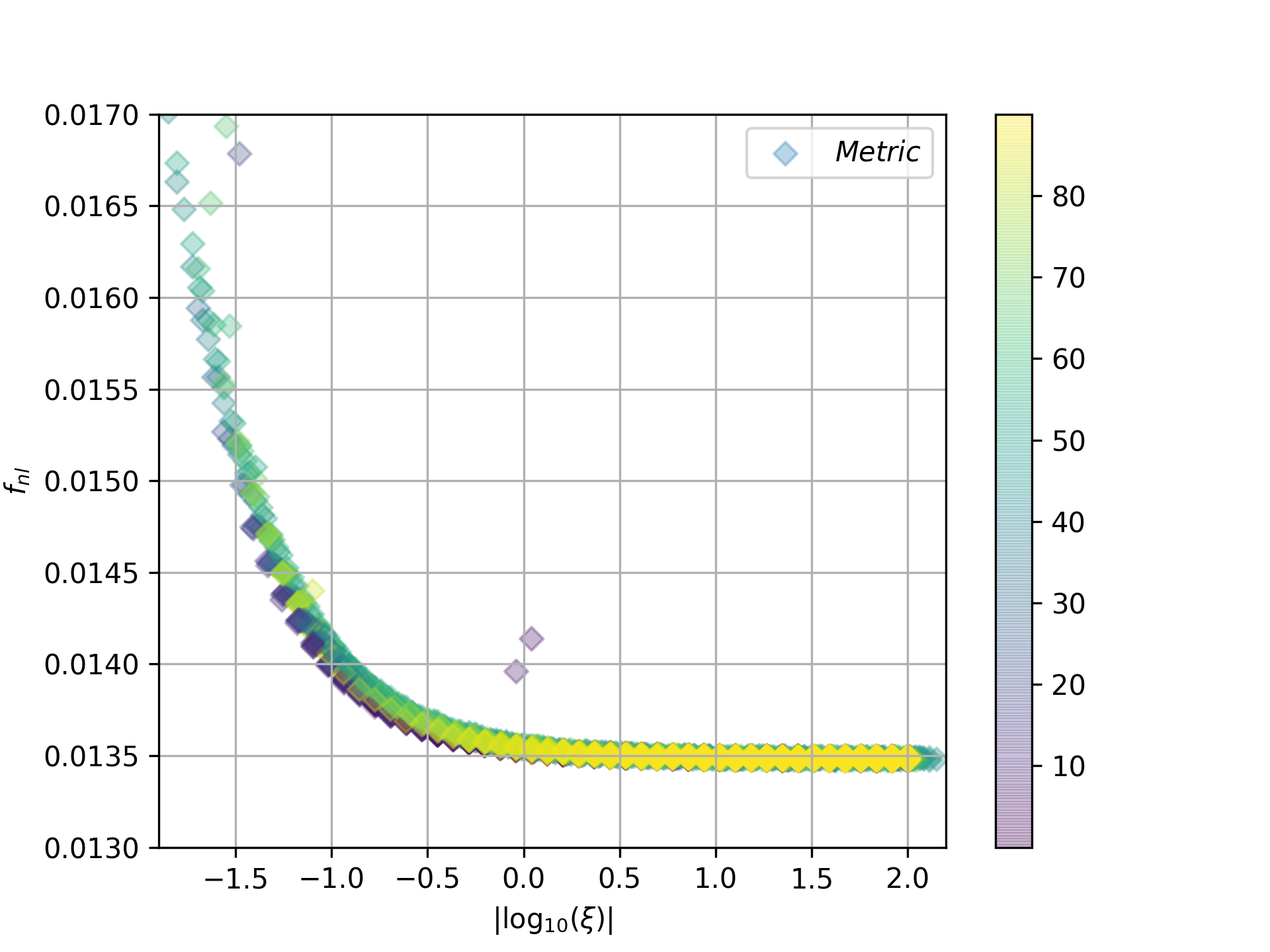}
		\includegraphics[width=0.45\textwidth]{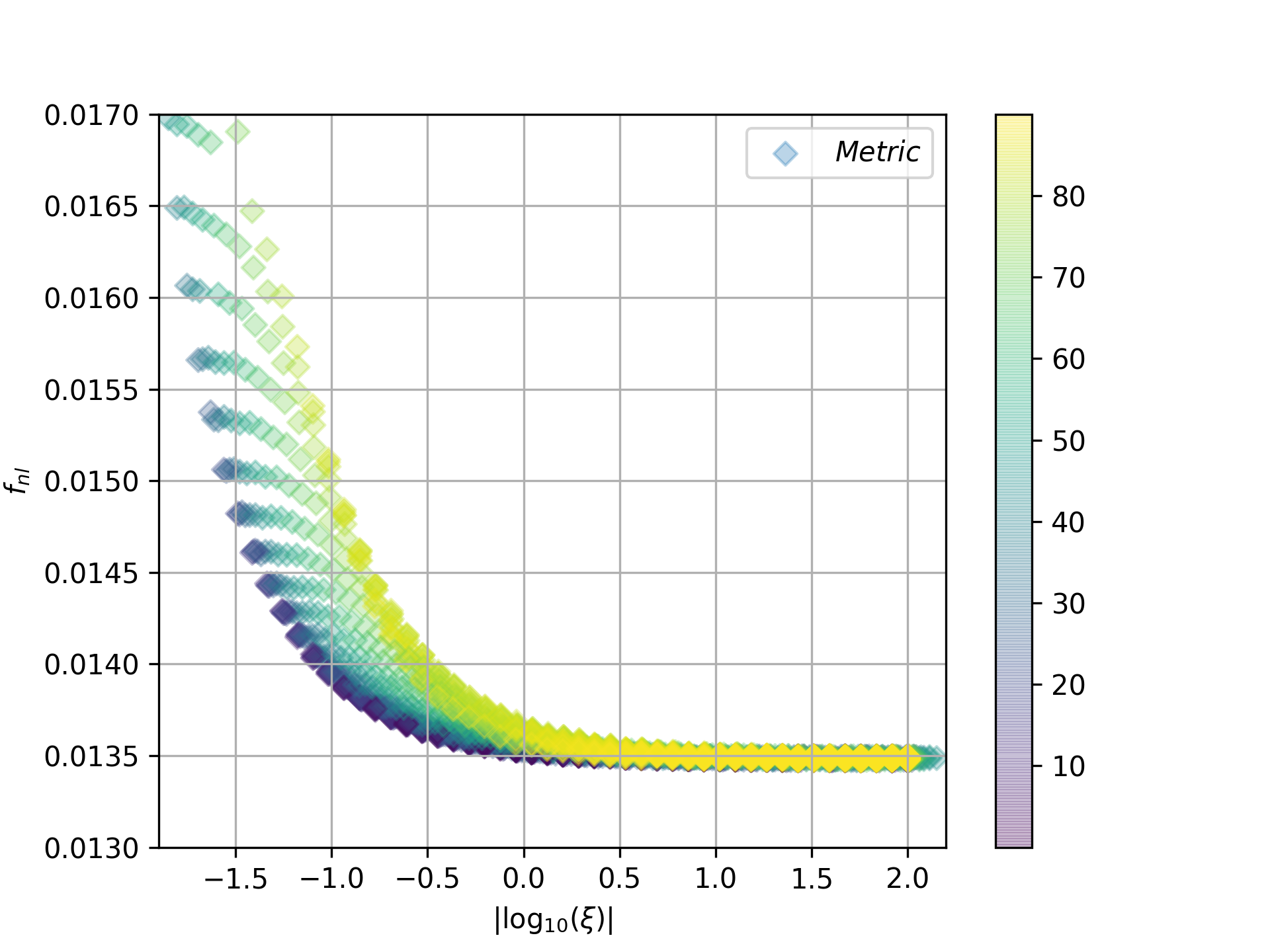}
	\caption{Predictions for $n_s$ (top) $r$ (middle) and $f_{\rm NL}$ (bottom) as a function of $\xi$ along the x-axis and $\theta=\tan^{-1}(\xi_\sigma/\xi_\phi)$ (illustrated by the color gradient in degrees) in metric gravity for $n=2$ and for the same $\lambda_\sigma/\lambda_\phi$ ratios as in Fig. \ref{fig:ICs}.}\label{vartheta}
\end{figure}

We see in Fig.~\ref{vartheta} that the results depend crucially on the ratio of the parameters in the potential, $\lambda_I$. When the parameters for both fields are similar, the observables quickly approach a single limiting value corresponding to the single-field case, while for the larger $\lambda_I$ ratio 
the predictions are substantially broadened throughout the entire $\xi$ range, with a clear dependence on the angular parameter $\theta$. The trajectories in $(n_s,r)$ space as a function of $\xi$ are also broadened, as is also clear in Fig.~\ref{nsr}. The predictions are thus somewhat different from the single field case for low and intermediate values of $\xi$, but converge to the same limit for sufficiently large $\xi$.

Having now analysed the dependence on both $\xi$ and $\theta$, we confirm that the results resemble the single-field case for both metric and Palatini gravity. The differences between single-field and multifield that do arise are apparent in the spread in the results for low values of $\xi$. This spread is due to a larger dependence on the initial conditions of the fields and on the direction in $\xi_I$ parameter space. At strong coupling, all the results found asymptote to the single-field ones. This similarity may be somewhat surprising, given that in the multifield case the field-space can be curved. We now show the reasons why this additional multifield effect is not affecting the results at strong coupling.

\begin{figure}
	\centering
	       \includegraphics[width=0.45\textwidth]{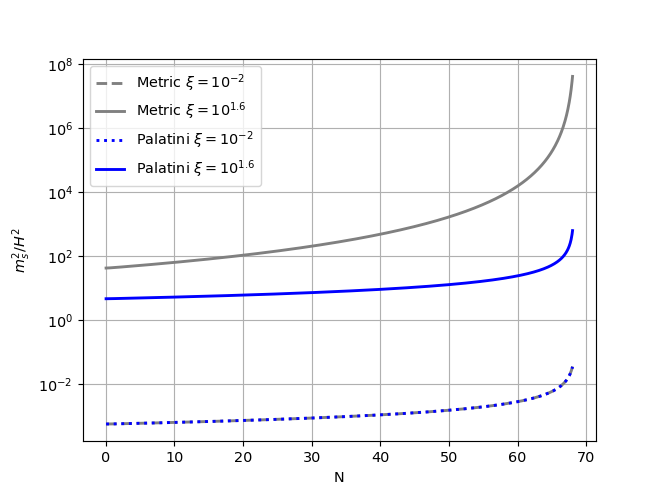}
	       \includegraphics[width=0.45\textwidth]{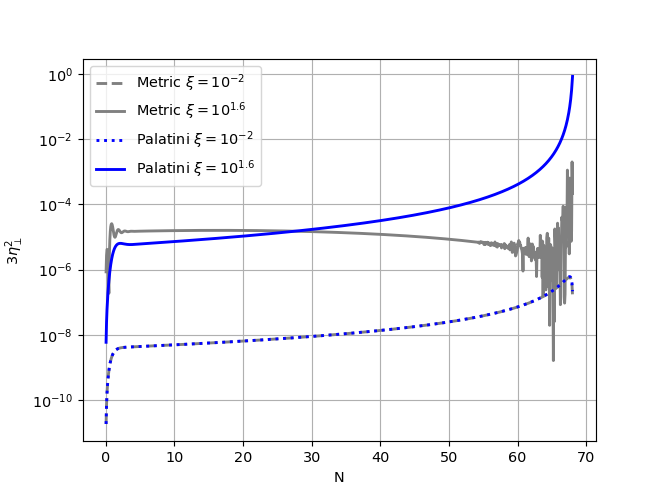}
        	\includegraphics[width=0.45\textwidth]{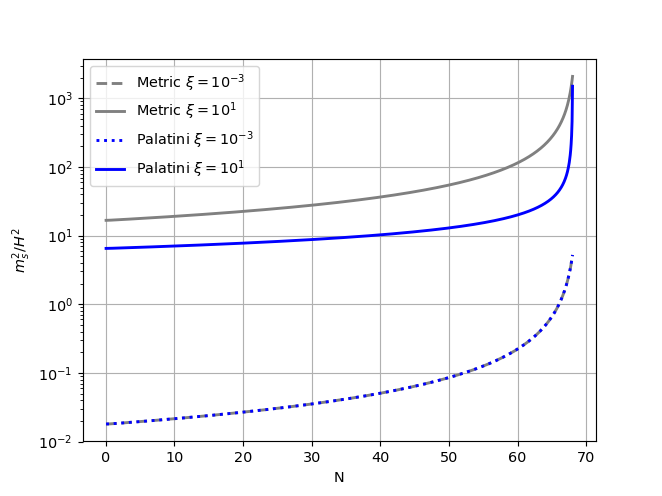}
        	\includegraphics[width=0.45\textwidth]{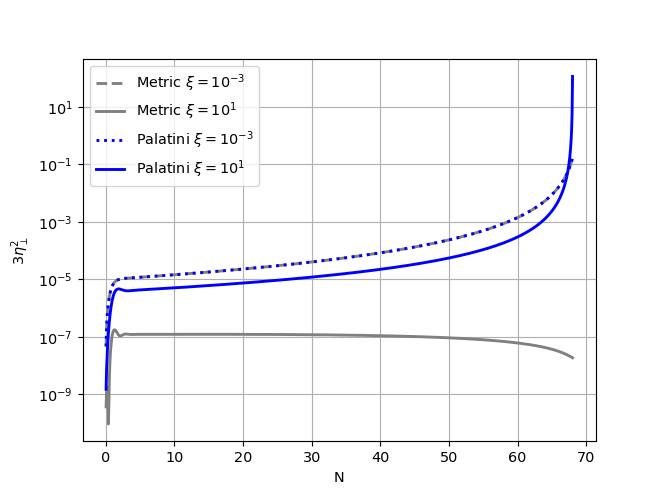}
		\includegraphics[width=0.45\textwidth]{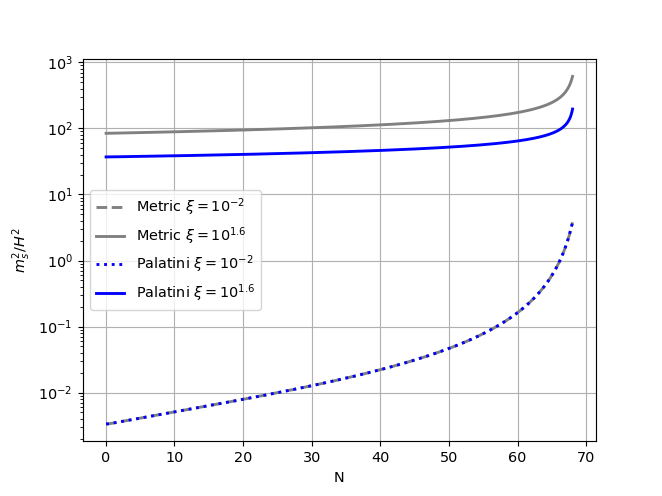}
	       \includegraphics[width=0.45\textwidth]{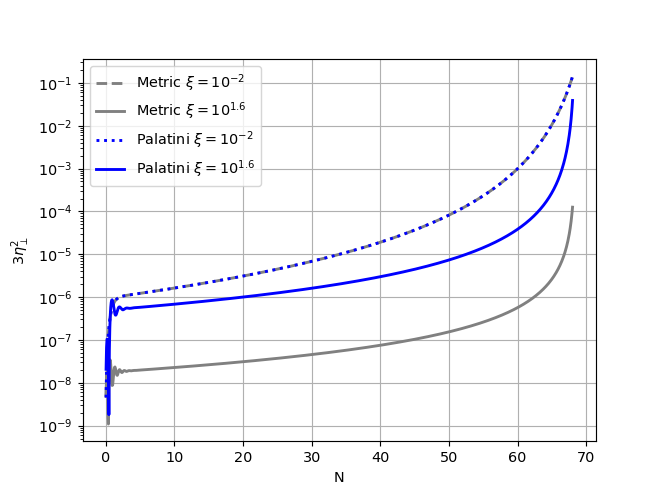}
        	\includegraphics[width=0.45\textwidth]{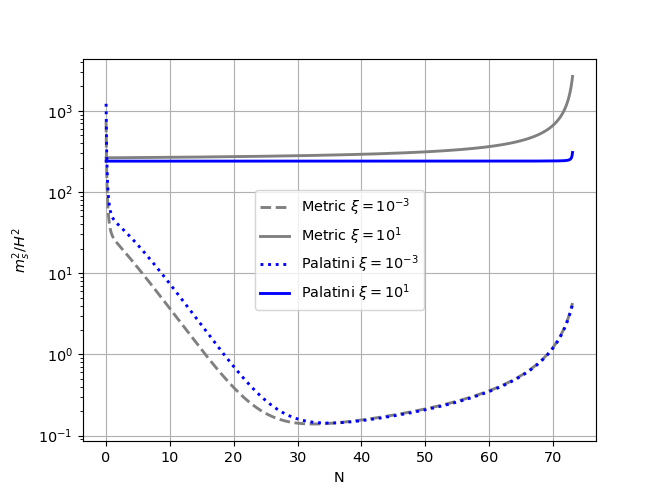}
        	\includegraphics[width=0.45\textwidth]{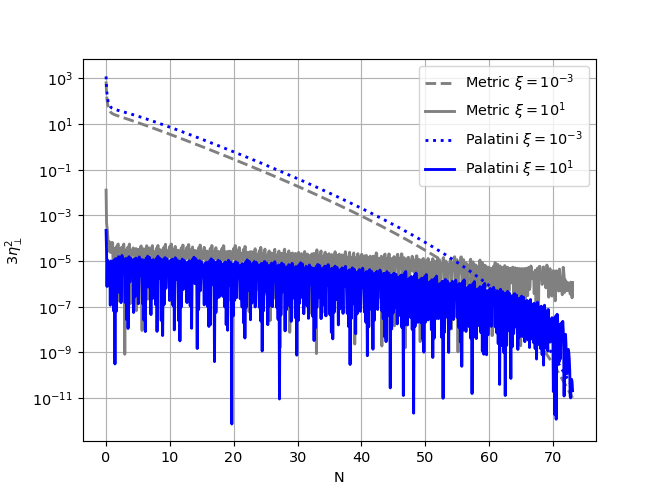}
	\caption{Evolution of the effective mass normalized to $H$ and bending parameter $\eta_\perp^2$ for metric (grey) and Palatini gravity (blue), $n=(1/2,1,3/2,2)$ from top to bottom. The dashed lines represent a sample with a small magnitude of the coupling parameters $\xi$ whereas the solid lines represents one with a large coupling. }\label{fig:EfMs}
\end{figure}

We first note that field-space curvature does not directly affect the evolution of the field fluctuations in the inflationary direction. This is because the field-space Riemann tensor appears in the effective mass matrix of the fluctuations, $m^I_L$, in the following term
\begin{equation}
m^I_L\supset R^I_{JKL} \dot{\phi^J}\dot{\phi^K}\,.
\end{equation}
To obtain the term relevant for the fluctuations in the inflationary direction, one must multiply $m^I_L$ with $\dot{\phi^L}$, which always results in zero for the term shown above, given the symmetries of the Riemann tensor. There is, however, an effect on the entropy perturbations, as they are sensitive to the perpendicular projection of the effective mass matrix. For the two-field case, the total effective mass for those fluctuations is given by
\begin{equation}
\label{eff-mass_eq}
\frac{m_s^2}{H^2}= \frac{U_{;ss}}{H^2}+3\eta_\perp^2+\epsilon R_{\text{fs}}\,,
\end{equation}
in which $\eta_\perp=U_{,s}/H\dot\phi$ is a measure of the bending of the trajectory, $\dot\phi=\sqrt{G_{IJ}\dot{\phi^I}\dot{\phi^J}}$, $s$ is the field coordinate in the entropic direction --- the direction perpendicular to $\dot{\phi^I}$ --- and $R_{\text{fs}}$ is the Ricci scalar of the field-space manifold. The effect of the curvature is somewhat less relevant if $R_{\text{fs}}$ is positive, as it simply contributes to a smaller amplitude of the  entropy perturbations. If it is negative, however, it reduces the effective mass and may even render it tachyonic should it be large enough~\cite{Renaux-Petel:2015mga}, thus dangerously enhancing the entropy fluctuations. Our numerical results seem to indicate that this never occurs, given their similarity with the single-field results, for which the curvature is not present. We can verify this by checking whether the condition $m_s^2>0$ is always verified in our numerical results. We can see this in Fig.~\ref{fig:EfMs}, in which we show that the effective mass is always positive for all values of $n$ studied above. When $\xi$ is large, the effective mass is also large, with the dominant contribution coming from the first term on the right hand side of Eq.~\eqref{eff-mass_eq}, the Hessian of the potential. Specifically, the effective mass values calculated in the metric and Palatini cases are equivalent for small $\xi$ and consequently the resulting observables ($n_s$, $r$ and $f_{\rm NL}$) are affected in similar ways in both cases. Where the observables deviate between the two cases, i.e. for large $\xi$, the effective masses also deviate with an overall larger effective mass in the metric case.

The evolution of the entropy modes is independent of the adiabatic modes on large scales, and thus only depends on the effective mass. They can, however, source curvature perturbations via the bending parameter $\eta_\perp$ in the equation \cite{Turzynski:2014tza,Wands:2000dp,Carrilho:2015cma}
\be
\label{dotzeta}
\dot\zeta\approx \sqrt{2}H\eta_\perp\frac{H}{M_{\rm P}\sqrt{\epsilon}}\frac{\delta S}{H}\,,
\ee
with $\delta S$ the fluctuations in the entropic direction. Thus, we can recover the single-field results if $\eta_\perp$ is sufficiently small. We can estimate the entropy fluctuations via their variance $\delta S \sim H^2/m_s$. Furthermore, we note that ${H}/\left({M_{\rm P}\sqrt{\epsilon}}\right)$ is approximately the value of $\zeta$ at horizon crossing, $\zeta_*$, and that the typical time scale associated to its variation is $H$, making $H\zeta_*$ the natural size of $\dot\zeta$, should it vary considerably. Given these arguments, we can rewrite Eq.~\eqref{dotzeta} as
\be
\label{dotzeta2}
\frac{\dot\zeta}{H\zeta_*} \sim \frac{\eta_\perp H}{m_s}\,,
\ee
and conclude that if the right-hand-side of Eq.~\eqref{dotzeta2} is much smaller than 1, the evolution of $\zeta$ is negligible. Therefore, to determine the importance of entropy fluctuations in the evolution of adiabatic ones, we must only calculate $\eta_\perp H/m_s$. In the right panel of Fig.~\ref{fig:EfMs}, we show the size of $\eta_\perp^2$ during inflation. Comparison with the effective mass shown in the left panel demonstrates that the bending parameter is sub-dominant relative to the effective mass. For example, for the $n=1$ metric case the ratio $\eta_\perp^2 H^2/m^2_s\sim 10^{-3}$ when $\xi$ is small and for large $\xi$, $\eta_\perp^2 H^2/m^2_s\sim 10^{-8}$, demonstrating that the entropy fluctuations are negligible at strong coupling. Comparing the metric and Palatini case for small $\xi$ we see that the results for the evolution of $\eta_\perp$ are the same. For large $\xi$, the evolutions diverge and $\eta_\perp$ in the metric case decays, while it grows in the Palatini case.

\subsection{Extension to scenarios with higher number of fields}
\label{3fieldcase}

We have also extended our calculations to the three-field case for $n_s$ and $r$. We found that the results resemble those for the two-field case, converging to the same limit in the strong coupling approximation for both metric and Palatini gravity. The main difference is the spread in observable space, which is substantially larger than in the two-field case. This is a consequence of the increased number of possible background field trajectories that result in successful inflation in higher field-space dimensions as well as the larger number of free parameters. This can affect the ability of distinguishing between different models, with some results for the Palatini model giving the same observables as those for the metric case, even at strong coupling for the latter. The strongly coupled Palatini case is still distinctive, given its very low tensor-to-scalar ratio prediction.

With an even larger number of fields, these predictions are expected to broaden further, but may ultimately converge again, in a statistical sense, 
as such a behaviour has been demonstrated in other scenarios with random potentials and very large numbers of fields~\cite{Aazami:2005jf,Easther:2005zr,Easther:2013rva,Dias:2016slx,Dias:2017gva,Bjorkmo:2017nzd}.

\section{Conclusions}
\label{conclusions}

We studied multifield inflation in scenarios where the fields are coupled non-minimally to gravity via $\xi_I(\phi^I)^n g^{\mu\nu}R_{\mu\nu}$. We concentrated on the so-called $\alpha$-attractor models with the potential $V=\lambda_I^{(2n)}  M_{\rm P}^{4-2n} (\phi^I)^{2n} $ in two formulations of gravity: in the usual metric case where $R_{\mu\nu}=R_{\mu\nu}(g_{\mu\nu})$, and in the Palatini formulation where also the connection $\Gamma$ and hence also $R_{\mu\nu}=R_{\mu\nu}(\Gamma)$ are independent variables. 

As the main result, we showed that the curvature of the field-space in the Einstein frame has no influence on the inflationary dynamics at the limit of large $\xi_I$, and one effectively retains the single-field case  regardless of the underlying theory of gravity. In the metric case this means that multifield models approach the single-field $\alpha$-attractor limit, whereas in the Palatini case the attractor behaviour is lost also in the case of multifield inflation.


\section*{Acknowledgements}
P.C. is supported by the Funda\c{c}\~{a}o para a Ci\^{e}ncia e Tecnologia (FCT) grant SFRH/BD /118740/2016, D.J.M. is supported by a Royal Society University Research Fellowship, J.W.R. is supported by a studentship jointly funded by Queen Mary University of London and by
the Frederick Perren Fund of the University of London and T.T. is supported by the U.K. Science and Technology Facilities Council grant ST/J001546/1.

\bibliography{palatini_2F}


\end{document}